\documentclass[fleqn,usenatbib]{mnras}
\DeclareRobustCommand{\DE}[3]{#2}
\DeclareRobustCommand{\DEL}[3]{#2}

\usepackage{newtxtext,newtxmath}
\usepackage[T1]{fontenc}
\usepackage{ae,aecompl}
\usepackage{comment}
\usepackage{caption}
\usepackage{subcaption}
\usepackage{graphicx}	% Including figure files
\usepackage{amsmath}	% Advanced maths commands
\graphicspath{ {images} }
\usepackage{xcolor,soul}

\definecolor{lightblue}{rgb}{.70,.95,1}
\sethlcolor{lightblue} 
\usepackage{xspace} % Define commands that appear not to eat spaces
\newcommand{\teff}{\ensuremath{T_{\mathrm{eff}}}\xspace}
\newcommand{\kms}{\ensuremath{\rm{km}\,s^{-1}}\xspace}
\newcommand{\logg}{\ensuremath{\log g}\xspace}
\newcommand{\feh}{\rm{[Fe/H]}\xspace}

\newcommand{\alphafe}{\rm{[\ensuremath{\alpha}/Fe]}\xspace}
\newcommand{\Gaia}{\textit{Gaia}\xspace}
\newcommand{\CaHK}{\emph{CaHK}\xspace}
\newcommand{\Pristine}{\emph{Pristine}\xspace}
\newcommand{\bprp}{\ensuremath{(\rm{BP}-\rm{RP})_0}\xspace}

\title[Pristine Sagittarius]{The Pristine Inner Galaxy Survey (PIGS) IV: A photometric metallicity analysis of the Sagittarius dwarf spheroidal galaxy\thanks{based on observations made with the Canada-France-Hawaii Telescope (CFHT) and the Anglo-Australian Telescope (AAT)}}

\author[S. Vitali et al.]{
Sara Vitali,$^{1}$\thanks{E-mail: sara.vitali@mail.udp.cl}
Anke Arentsen,$^{2}$
Else Starkenburg,$^{3}$
Paula Jofr\'e,$^{1}$
Nicolas F. Martin,$^{2,4}$
\newauthor
David S. Aguado,$^{5,6}$
Raymond Carlberg,$^{7}$
Jonay I. Gonz\'alez Hern\'andez,$^{8,9}$
Rodrigo Ibata,$^{2}$
\newauthor
Georges Kordopatis,$^{10}$
Khyati Malhan,$^{4}$
Pau Ramos,$^{2}$
Federico Sestito,$^{11}$
Zhen Yuan,$^{2}$
Sven Buder,$^{12,13}$
\newauthor
Geraint F. Lewis,$^{14}$
Zhen Wan,$^{15}$
Daniel B. Zucker$^{16,17}$
\newauthor
\\
\\
$^{1}$ N\'ucleo de Astronom\'ia \& Millenium Nucleus ERIS, Facultad de Ingenier\'ia y Ciencias Universidad Diego Portales, Ej\'ercito 441, Santiago, Chile\\
$^{2}$Universit\'e de Strasbourg, CNRS, Observatoire astronomique de Strasbourg, UMR 7550, F-67000 Strasbourg, France\\
$^{3}$Kapteyn Astronomical Institute, University of Groningen, Postbus 800, 9700 AV, Groningen, the Netherlands\\
$^{4}$Max-Planck-Institut f\"ur Astronomie, K\"onigstuhl 17, D-69117 Heidelberg, Germany\\
$^{5}$Dipartimento di Fisica e Astronomia, Universitá degli Studi di Firenze, Via G. Sansone 1, I-50019 Sesto Fiorentino, Italy\\ 
$^{6}$ INAF/Osservatorio Astrofisico di Arcetri, Largo E. Fermi 5, I-50125 Firenze, Italy\\ 
$^{7}$Department of Astronomy \& Astrophysics, University of Toronto, Toronto, ON M5S 3H4, Canada\\
$^{8}$Instituto de Astrofísica de Canarias (IAC), E-38200 La Laguna, Tenerife, Spain\\
$^{9}$Universidad de La Laguna, Dept. Astrofísica, E-38206 La Laguna, Tenerife, Spain\\
$^{10}$Universit\'e C\^ote d'Azur, Observatoire de la C\^ote d'Azur, CNRS, Laboratoire Lagrange, Nice, France\\
$^{11}$Department of Physics and Astronomy, University of Victoria, PO Box 3055, STN CSC, Victoria BC V8W 3P6, Canada\\
$^{12}$Research School of Astronomy \& Astrophysics, Australian National University, ACT 2611, Australia\\
$^{13}$Center of Excellence for Astrophysics in Three Dimensions (ASTRO-3D), Australia\\
$^{14}$Sydney Institute for Astronomy, School of Physics, A28, The University of Sydney, NSW 2006, Australia\\
$^{15}$School of Astronomy and Space Science, University of Science and Technology of China, Hefei, Anhui 230026, China\\
$^{16}$Department of Physics and Astronomy, Macquarie University, Sydney, NSW 2109, Australia\\
$^{17}$Macquarie University Research Centre for Astronomy, Astrophysics \& Astrophotonics, Sydney, NSW 2109, Australia\\
}

\date{Accepted 20 September 2022. Received 3 October 2022, in original form 26 April 2022}

\pubyear{2022}

\begin{document}
\label{firstpage}
\pagerange{\pageref{firstpage}--\pageref{lastpage}}
\maketitle

\begin{abstract}
We present a comprehensive metallicity analysis of the Sagittarius dwarf spheroidal galaxy (Sgr dSph) using \Pristine \CaHK photometry. We base our member selection on \Gaia EDR3 astrometry applying a magnitude limit at $G_{0} = 17.3$, and our population study on the metallicity-sensitive photometry from the \Pristine Inner Galaxy Survey (PIGS). Working with photometric metallicities instead of spectroscopic metallicities allows us to cover an unprecedented large area ($\sim 100$ square degrees) of the dwarf galaxy, and to study the spatial distribution of its members as function of metallicity with little selection effects. Our study compares the spatial distributions of a metal-poor population of 9719 stars with \feh $< -1.3$ and a metal rich one of 30115 stars with \feh $> -1.0$. The photometric Sgr sample also allows us to assemble the largest sample of 1150 very metal-poor Sgr candidates (\feh $< -2.0$). By investigating and fitting the spatial properties of the metal-rich and metal-poor population, we find a negative metallicity gradient which extends up to 12 degrees from the Sgr center (or $\sim 5.5$ kpc at the distance of Sgr), the limit of our footprint. We conclude that the relative number of metal-poor stars increases in the outer areas of the galaxy, while the central region is dominated by metal-rich stars. These finding suggest an outside-in formation process and are an indication of the extended formation history of Sgr, which has been affected by the tidal interaction between Sgr and the Milky Way.

\end{abstract}


\section{Introduction}\label{intro}

The variety of galaxies present in the Universe, with their different shapes, features and sizes, suggests the existence of several formation processes behind galactic structures. Due to the mutual gravitational attractions, mergers have led to the formation of bigger structures. It has been widely acknowledged that satellite galaxies have been accreting onto the Milky Way (MW) \citep[see for a review e.g.][]{2005ApJ...635..931B,2016ARA&A..54..529B}. The dwarf galaxies that orbit around the Milky Way are insightful laboratories to learn about the early evolution of our Galaxy, as they are relics of the main building blocks of the Galactic halo \citep{2012ApJ...759..115F}. 

The Sagittarius dwarf spheroidal galaxy (Sgr dSph) is one of such galaxies, discovered by \citet{1994Natur.370..194I} in the direction of the Galactic bulge. It is one of the biggest dwarf galaxies known around the Milky Way, with an estimated stellar mass of $ \sim 4.8 \times 10^{8}\mathrm{M}_{\odot}$ \citep{2020MNRAS.497.4162V}, and among the most luminous with $M_{\mathrm{v}} \sim -15.1/-15.5$ \citep{2010ApJ...712..516N}. Its core is located on the opposite side of the Galactic centre, at a relatively nearby heliocentric distance of $\approx 26.5\,\mathrm{kpc}$ \citep{2004MNRAS.349.1278M, 2020MNRAS.495.4124F, 2020MNRAS.497.4162V}. 
Sgr is a compelling example of an on-going merger with the Milky Way, in which the system is being disrupted by the tidal interaction with our Galaxy \citep{1997hst..prop.7463I,1998ApJ...508L..55M,2014MNRAS.437..116B}, with the first in-fall occurring about 5 Gyr ago \citep{2020NatAs...4..965R}. Many of its stars have been stripped away from the core in long tidal streams \citep{2001ApJ...551..294I,2003ApJ...599.1082M,2010ApJ...714..229L} that wrap around the Milky Way. Despite its inevitable destruction, the core of the Sgr dwarf galaxy is still visible. However, the projected proximity to the Galactic bulge has made the study of the dSph galaxy challenging due to the contamination from Milky Way foreground stars and extinction by interstellar dust. 

With its history of tidal disruption, Sgr is a unique workshop for examining the physical aspects connected to the chemical evolution from the perspective of the hierarchical galaxy formation scenario. 
In recent years, a number of studies have been dedicated to disentangle the complex Sgr star formation history (SFH), based either on high-resolution spectroscopy \citep[e.g.][]{2000A&A...359..663B, 2002astro.ph..5411S, 2005A&A...437..905S, 2005A&A...441..141M,2007ApJ...670..346C,2013ApJ...778..149M, 2017ApJ...845..162H,2018ApJ...855...83H} or photometric techniques \citep{1999MNRAS.307..619B, 2000AJ....119.1760L, 2007ApJ...667L..57S}. They found that Sgr has experienced many bursts of star formation
that resulted in stellar populations with different ages and metallicities. These are described in detail for instance in the work of \citet{2007ApJ...667L..57S}, in which the monotonically varying age-metallicity distribution consists of four different populations: a dominant intermediate-age stellar population aged $\sim 4-8 $ Gyr with $-0.6 \lesssim \feh \lesssim -0.4$, a younger and more metal-rich stellar population of $2-3$ Gyr old with $\feh \sim -0.1$, a small population younger than 2 Gyr with super-solar metallicities ($\feh \sim +0.5$), and a metal-poor population with $\feh \sim -1.2$ and ages $>10$ Gyr.

Sgr is one of the most massive satellite galaxies around the MW, after the Large Magellanic Cloud and Small Magellanic Cloud. The stellar mass-metallicity relation for dwarf galaxies predicts that more massive galaxies show higher average metallicity \citep{2013ApJ...779..102K}. The predominance of a relatively metal-rich population in the Sgr core (with the bulk of the stars having an average $\feh \sim -0.5$, \citealt{2003ApJ...597L..25M,2007ApJ...667L..57S,2017A&A...605A..46M}) makes the identification and study of metal poor stars particularly difficult. Sgr does also host an old and metal-poor component \citep[$\feh < -1.0$ and age $\sim 10$ Gyr,][]{2003ApJ...597L..25M, 2007ApJ...667L..57S,2008AJ....136.1147B}, but to date, only $\sim$20 very metal-poor (VMP, $\feh < -2.0$) Sgr stars have been discovered and studied with either high- or low-resolution spectroscopy \citep{2008AJ....136.1147B, 2017A&A...605A..46M, 2018ApJ...855...83H, 2019ApJ...875..112C, 2020ApJ...901..164C}. This very metal-poor population in Sgr has important implications when studying galaxy evolution. They are archaeological fossils from the earliest time which will unveil the primitive stellar populations in the Sgr dwarf galaxy. One theoretical expectation is that smaller dwarf systems may have contributed to the formation of more massive ones, which could have happened to Sagittarius \citep{2020ApJ...901..164C}. A recent work by \citet{2022ApJ...926..107M} found that the metal-poor Elqui stream is associated with Sagittarius and was likely accreted inside the Sgr dSph.

\begin{figure}
\centering
\includegraphics[width=1.0\hsize]{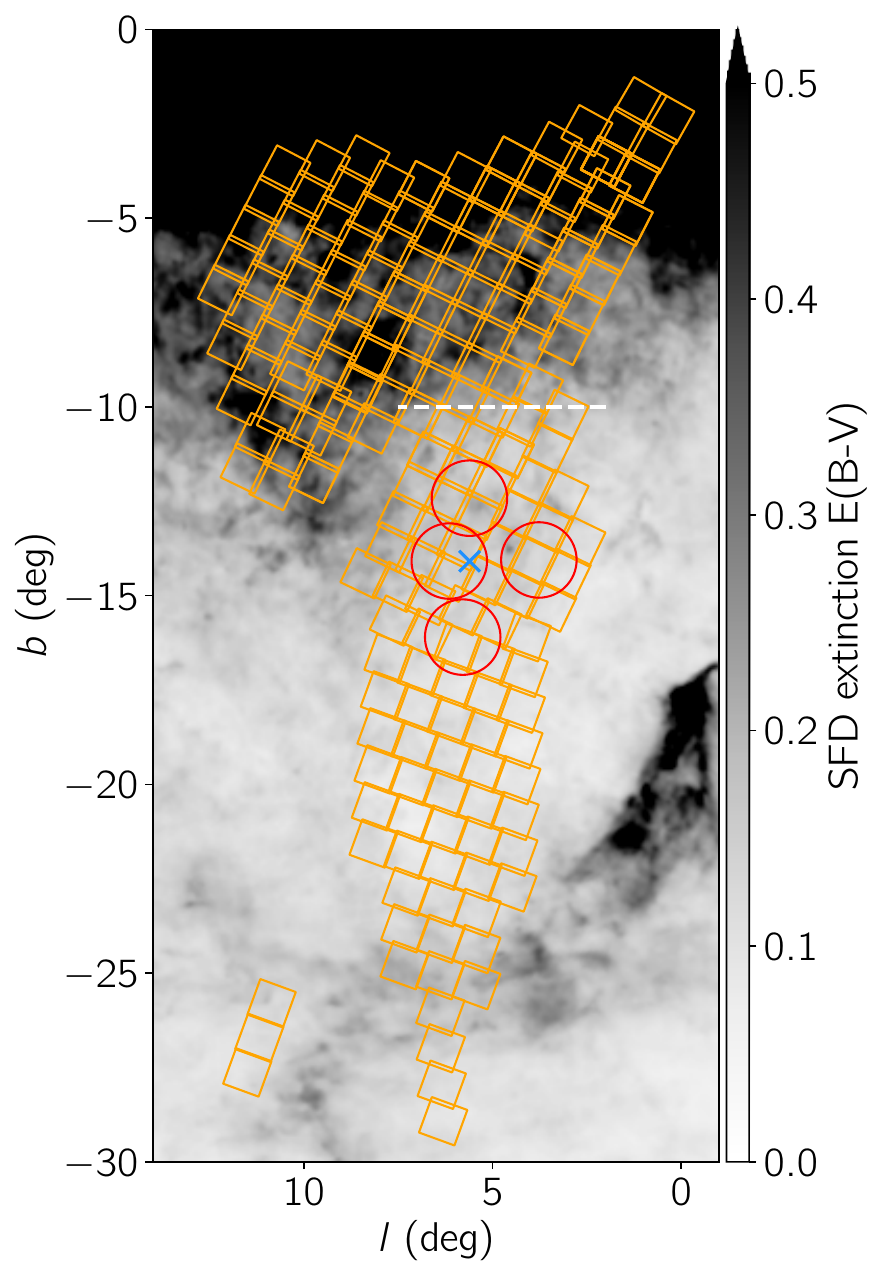}
\caption{The photometric fields of the Southern PIGS photometric footprint are shown in orange, with the Sgr extension present below the white dashed line at $b=-10$. The four red circles indicate fields with dedicated Sgr spectroscopic follow-up in PIGS. The extinction map is from \citet{1998ApJ...500..525S}, where for the sake of contrast we fix the upper limit of the colour bar at 0.5. The location of M54 has been highlighted with a blue cross.
}
\label{fig:footprint} 
\end{figure}

Studying the spatial distribution of different stellar populations is key, because it helps us to understand the various episodes of star formation which have occurred during the evolution of Sgr.  
The correlation of the present spatial distributions of populations of different metallicities and ages provides hints about the primitive distribution of the gas from which they formed. Using chemical abundances of a sample of Sgr stars, \citet{2017A&A...605A..46M} revealed a metallicity gradient inside the core of the dwarf galaxy, supporting the hypothesis of a complex SFH. How the evolution of the galaxy has affected the spatial distribution of the different stellar populations is still an open debate, which needs more extended and comprehensive samples, and especially for the more metal-poor component. For this purpose, one would ideally have a large, homogeneous and clean sample of Sgr stars with available metallicities.

The incredible data collected by the Gaia mission, and especially the arrival of the high-accuracy \textit{Gaia} EDR3 astrometry \citep{2021A&A...649A...1G}, allow for the building of a robust sample of Sgr member stars. Relying on photometric metallicities instead of spectroscopic metallicities allows the use of a much larger and more homogeneous sample to investigate the global metallicity structure of the galaxy. 
In this context, a great data set is the photometric \Pristine Inner Galaxy Survey \citep[PIGS,][]{arentsen2020pristine}, consisting of metallicity-sensitive \CaHK photometry of stars in the Milky Way bulge region. PIGS includes a region focused on Sgr, from the highest-density area of the system until the onset of its tidal stream. 

This paper is devoted to a photometric metallicity analysis of the Sgr galaxy, carried out thanks to the combination of the \textit{Gaia} astrometry and broad-band photometry and the PIGS metallicity-sensitive \CaHK photometry which cover about 100 square degrees of Sgr. The data and the member selection are presented in Section~\ref{sect:data}. The combination of \Gaia and \Pristine leads to an unprecedented large Sgr sample of 44785 stars with $G_0 < 17.3$ with available metallicity information, enabling a wide investigation of the spatial distributions of the stellar populations with different metallicities hosted in Sgr. We analysed the different spatial distributions of populations of different metallicities ($\feh < -2.0, < -1.3$ and $> -1.0$) in Section~\ref{met_analysis}, and present a large sample of 1150 new very metal-poor candidates. We study the metallicity gradient, which extends to at least $\sim 12^{\circ}$ from the centre of the Sgr remnant. Our approach is very effective in characterising the metallicity structure of Sgr. We discuss what our results can teach us about the (early) evolution of the Sagittarius system in Section~\ref{discussion}, and finish with conclusions and a discussion of future prospects in Section~\ref{conclusions}. 
\section{Data}\label{sect:data}
\subsection{Photometry}
In this paper we use the PIGS  photometry in the Sgr region, see Figure~\ref{fig:footprint}. The \Pristine survey (of which PIGS is an extension) has been ongoing since 2016 \citep{2017MNRAS.471.2587S} and has the main goal to search for and study the most metal-poor stars in and around the Milky Way. It makes use of the narrow-band \CaHK filter designed for MegaCam mounted on the Canada-France-Hawaii Telescope (CFHT), which covers the Ca II H\&K absorption lines that are sensitive probes of stellar metallicity. The targeting of metal-poor stars in the main \Pristine halo survey has been extremely efficient \citep{youakim17, aguado19}, and this efficiency has been demonstrated in the inner Galaxy as well \citep{arentsen2020pristine}. The Sgr extension of PIGS has not been used before, and we present it here for the first time. 

\subsubsection{Photometric field-to-field calibration}

The data reduction proceeds as in \citet{2017MNRAS.471.2587S} until the field-to-field calibration step. Figure~\ref{fig:footprint} presents each of the observed \CaHK MegaCam images in our footprint (orange squares), which may have slightly different zero-points of the order of a few tenths of a magnitude. In the main \Pristine halo survey, fields were calibrated with respect to each other by determining \CaHK offsets using the red part of the stellar locus of a \CaHK--SDSS colour-colour diagram (see Figure~7 in \citealt{2017MNRAS.471.2587S}). This part of the locus mostly consists of nearby dwarf stars. For this calibration, the photometry needs to be dereddened, which was done using the same reddening map for the halo survey, i.e. the Schlegel dust map, hereafter SFD \citep{1998ApJ...500..525S} map. This map contains the integrated reddening along the line of sight, which may be different than the actual reddening to the nearby dwarfs used for the calibration. Towards the halo, this difference is small and has not been taken into account. Towards Sgr, however, the reddening could be expected to change significantly as function of distance to the stars because we are looking through a large part of the disc, and we should not use the integrated SFD reddening. We therefore devised a new field-to-field calibration strategy specifically for the PIGS Sgr footprint ($b < -10$ degree). 

Instead of using only the red part of the stellar locus, we use the full \CaHK--Gaia stellar locus of nearby dwarfs and simultaneously fit for the \CaHK offset (shifting the locus by a constant) and the average foreground E(B$-$V) for each field (changing the shape of the locus). We select dwarf stars ($M_G<4$) with \CaHK uncertainties $<0.15$~mag, parallax uncertainties $< 20$\% and distances roughly between $500 - 1000$~pc (from the \Gaia parallaxes) as to not span too large a range in distances. We apply additional quality cuts to the astrometry (RUWE $< 1.2$, this is stricter than the cut we apply in Section~\ref{selection}, because we aim for higher quality for the calibration) and the photometry (the same cuts on BP$-$RP excess factor and variability as discussed in Section~\ref{selection}) to get to the final calibration sample. \begin{figure}
\centering
\includegraphics[width=0.8\hsize,trim={2.0cm 0.0cm 0.0cm 0.0cm}]{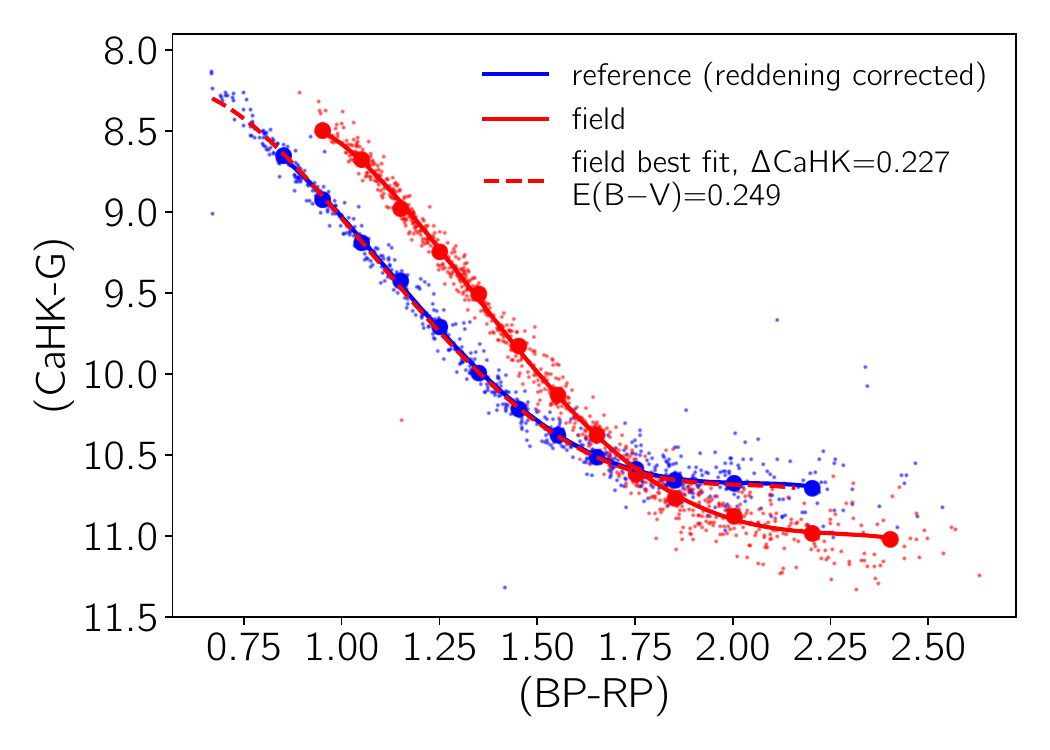}
\includegraphics[width=0.85\hsize,trim={0.0cm 0.0cm 0.0cm 0.0cm}]{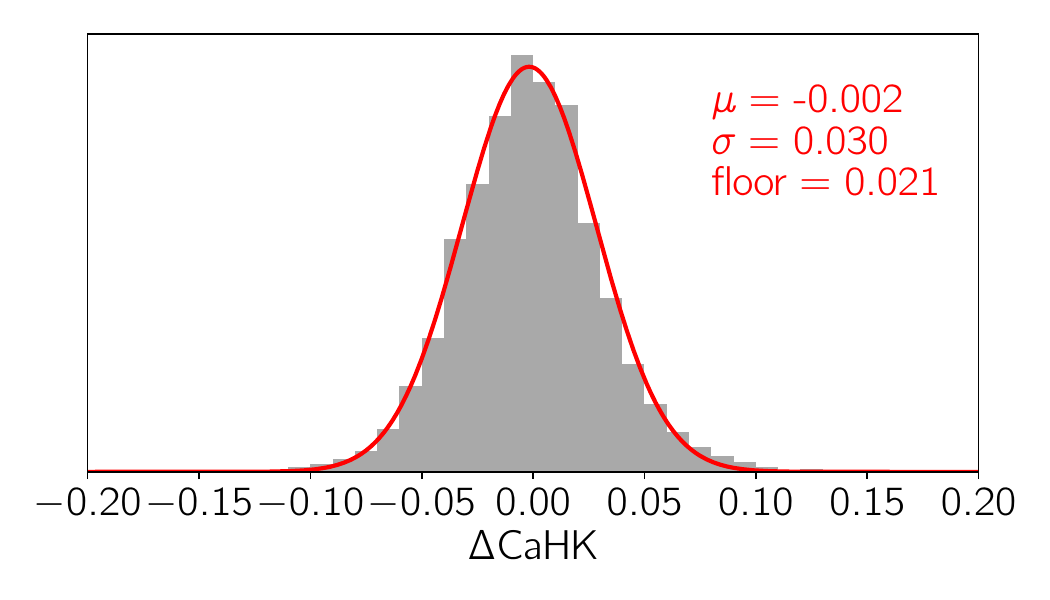}
\caption{Field-to-field calibration of the PIGS-Sgr footprint. Top panel: stellar locus used for the calibration, with the reference field in blue and an example field to calibrate in red. Large points indicate the median $(\CaHK-G)$ values in bins of $(\mathrm{BP}-\mathrm{RP})$, the solid lines are 4th order polynomials fitted through the medians. The red-dotted line indicates the best fit, with the best $\Delta\CaHK$ and E(B$-$V) indicated. Bottom panel: difference in \CaHK for all pairs of observations of a given star on two different MegaCam fields, for all fields in the Sgr footprint. The mean $\mu$, standard deviation $\sigma$, and uncertainty floor $\sigma/\sqrt{2}$ are indicated.}
\label{fig:calib} 
\end{figure}

We select a field with relatively low E(B$-$V)$_\mathrm{SFD}$ at $(l,b) = (6.25^{\circ}, -21.2^{\circ})$ as the reference field, and correct it for extinction using the integrated SFD map (which is likely not too bad 20 degrees away from the Galactic plane) with the filter coefficients as described in the next sub-section. We take the median $(\CaHK-G)_0$ in bins in \bprp of 0.1 (or 0.2 in the reddest part), requiring at least 5 stars per bin, and fit a 4th order polynomial to it. We do the same in each field to be calibrated, but for the reddened colours. We then compute the $\chi^2$ for a grid of $\Delta\CaHK$ and E(B$-$V) and adopt the best solution. We show the stellar locus of the reference field and an example field with its best fit in the top panel of Figure~\ref{fig:calib}. 

We test the quality of our calibration by determining the difference in \CaHK for all stars in the Sgr footprint observed on two different images, possible thanks to the overlap between fields. Taking only stars with \CaHK uncertainties less than 0.01, we infer a dispersion of 0.03 mag implying an uncertainty floor of 0.021 (dividing the dispersion by $\sqrt{2}$), see the bottom panel of Figure~\ref{fig:calib}. This is comparable to the uncertainty floor estimated for the main survey (0.02~mag, \citealt{2017MNRAS.471.2587S}), and corresponds to a metallicity uncertainty of $\sim0.15$~dex in our methodology (see Section~\ref{met_analysis}).

\subsubsection{Catalogue}
We cross-match the PIGS photometry with the astrometric data from \textit{Gaia} EDR3 \citep{2021A&A...649A...1G}, from which we use parallaxes and proper motions to perform a selection to isolate the member of Sgr (see Section~\ref{selection}). We also use the \textit{Gaia} broad-band photometry combined with the PIGS photometry to select stars of different metallicities (see Section~\ref{division}). Throughout this paper, we only use data cross-matched with the PIGS+\textit{Gaia} catalogue for Sgr. 

We correct for dust extinction using the SFD map. We use the colour-dependent extinction coefficients for the \Gaia EDR3 filters from \citet{casagrande2020effective} (adopting the ``FSF'' extinction law) and 3.924 for \CaHK \citep{2017MNRAS.471.2587S}.

\subsection{Member selection} \label{selection}

Establishing proper criteria to define the membership of stars to a dwarf galaxy is a crucial and challenging step. 
Sgr is highly affected by the contamination coming from different Galactic populations, namely the Galactic bulge, the thick disk and the thin disk \citep{1997AJ....113..634I}, and also the Galactic halo. It is necessary to find an accurate balance between reducing Milky Way contamination and not loosing too many Sgr candidates.

We apply the following cuts to obtain Sgr members from our PIGS+\textit{Gaia} catalogue. We only consider the part of the PIGS footprint with $ b < -10^{\circ}$, to avoid the region close to the Galactic plane which will have more contamination. We isolate Sgr candidates using a cut on the parallax to remove foreground stars ($|\varpi| \leq 2\,\epsilon_{\varpi}$). We limit our analysis to stars with $ \mathit{G_{0}}\leq 17.3$ to avoid fainter stars for which the uncertainties on the \Gaia astrometry and the \textit{CaHK} photometry become too large. 
Additionally, helium-burning (red clump and horizontal branch) stars start to contribute significantly at fainter magnitudes. Deriving photometric metallicities for the (bluer) horizontal branch is more difficult since the metallicity sensitivity of \Pristine reduces for hotter stars, our spectroscopic training sample (see the next section) does not include horizontal branch stars and some of the stars on the horizontal branch are photometrically variable. Excluding these bluer (more metal-poor) stars while including the (more metal-rich) red clump stars would bias our sample against metal-poor stars. This is another reason to limit the analysis to brighter stars only. This leads to a sample of $\sim 52000$ stars. We apply additional quality cuts on \texttt{RUWE} ($< 1.4$), \texttt{astrometric\_excess\_noise\_sig} ($\leq 2$) as described in \citet{2021A&A...649A...2L}, and the fidelity flag ($ > 0.5$) from \citet{2022MNRAS.510.2597R}. 

\begin{figure}
\includegraphics[width=0.45\textwidth]{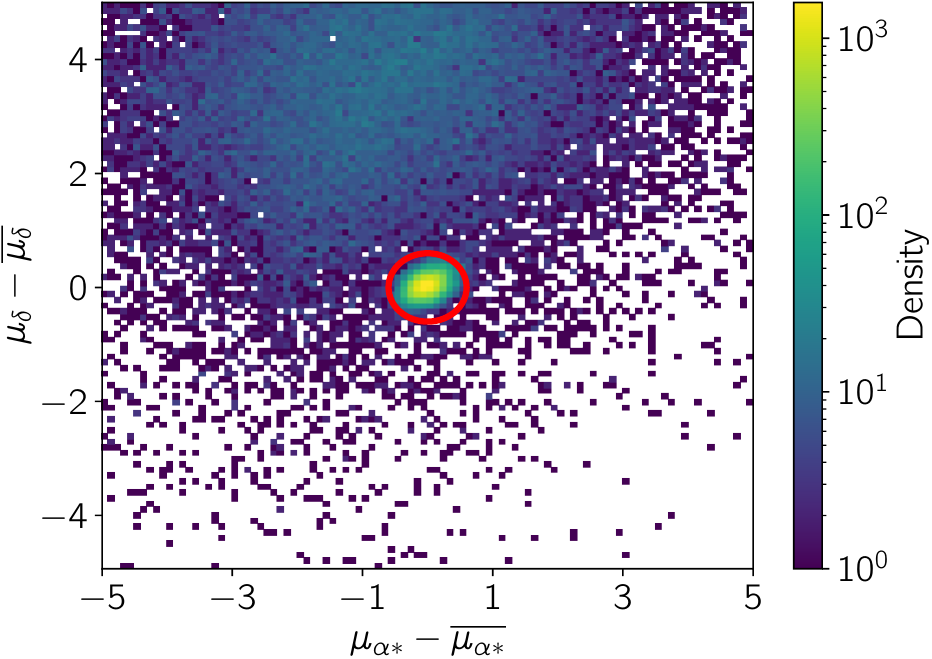}
\caption{Zoom in the distribution of the proper motion values derived from \textit{Gaia} EDR3 \citep{2021A&A...649A...1G} for the PIGS data-set after having isolated the Sgr candidates. These PM values are subtracted from the mean PM of the Sgr members ($\overline{\mu_{\alpha}}$) following expression (4) from \citet{2020MNRAS.497.4162V}. The Sgr galaxy is visible as a clear over-density centred at 0. We select Sgr stars within a radius of 0.6 mas/yr$^{-1}$, delimited by the red circle.}
\label{density}  
\end{figure}

\begin{figure}
     \centering\begin{subfigure}[b]{0.42\textwidth}
         \centering
         \includegraphics[width=\textwidth]{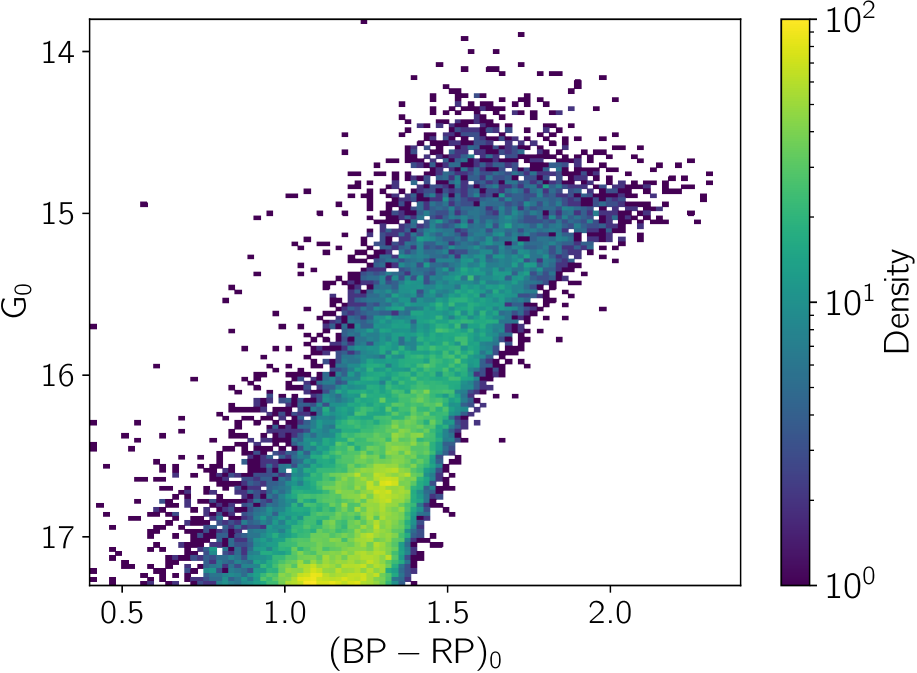}
     \end{subfigure}
     \hfill
     \begin{subfigure}[b]{0.42\textwidth}
         \centering
         \includegraphics[width=\textwidth]{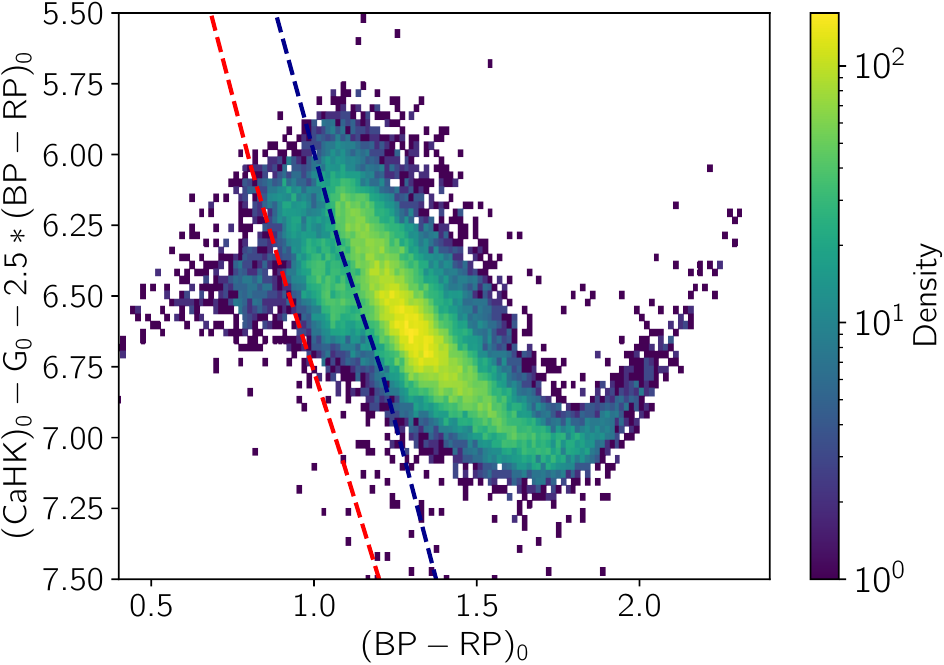}
     \end{subfigure}
     \caption{\textbf{Top:} Colour-magnitude diagram with \Gaia photometry for the Sgr selected members after applying all the cuts described in the text. \textbf{Bottom:} Colour-colour diagram of the same Sgr selection. The y-axis includes the metallicity-dependent \CaHK term. We identified stars below the red dashed line are as MW contamination, and removed them from the Sgr sample. The blue line delimits the Sgr helium-burning red clump sequence from the normal giants. }
\label{panel_2}
\end{figure}
\begin{figure*}
     \begin{subfigure}[b]{0.38\textwidth}
         \centering
         \includegraphics[width=\textwidth]{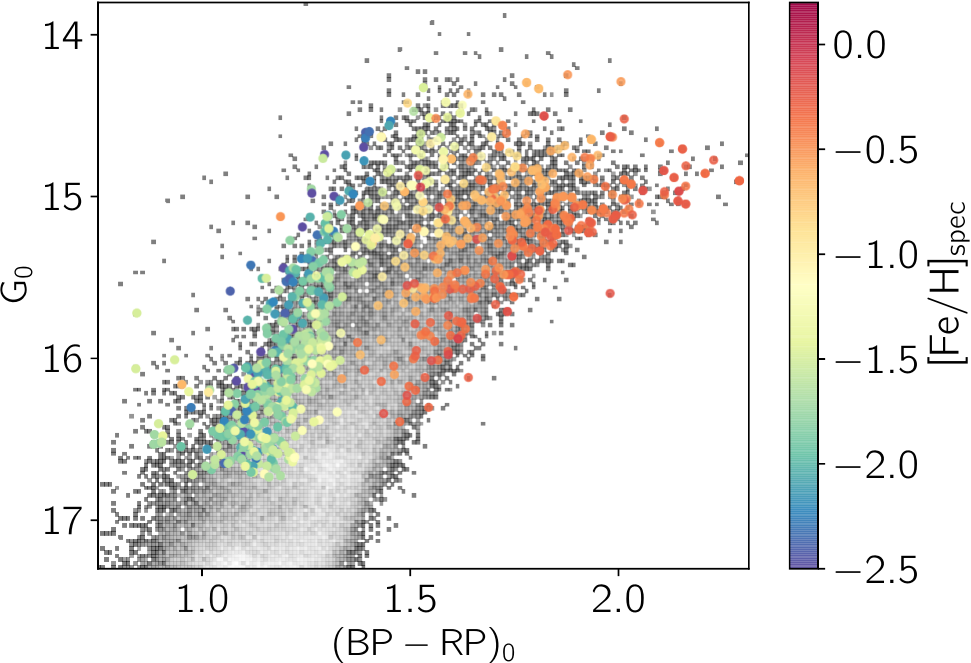}
     \end{subfigure}
     \hspace{0.15cm}
     \begin{subfigure}[b]{0.26\textwidth}
         \centering
         \includegraphics[width=\textwidth]{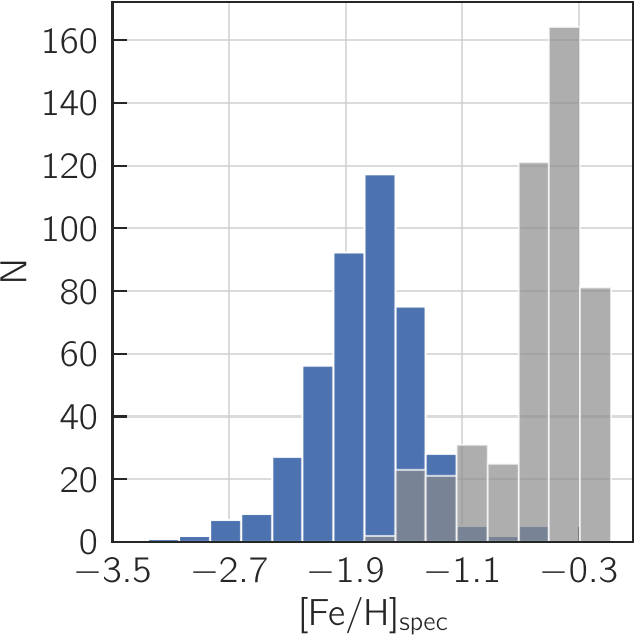}
     \end{subfigure}
     \hspace{0.1cm}
     \begin{subfigure}[b]{0.26\textwidth}
         \centering
         \includegraphics[width=\textwidth]{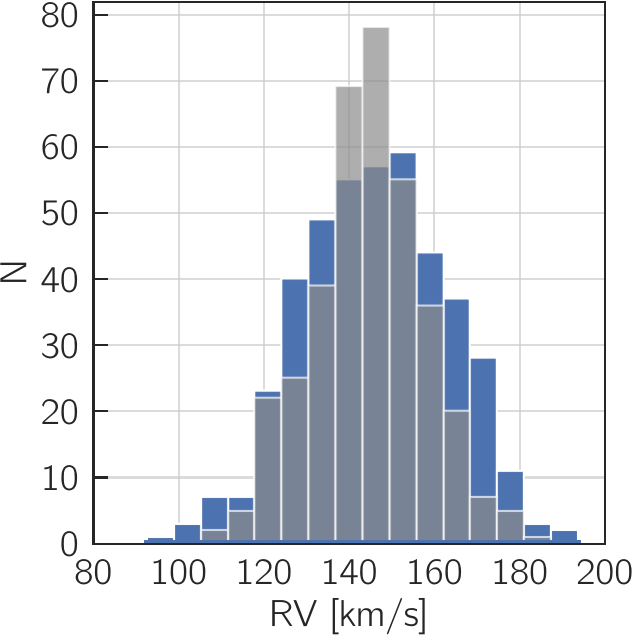}
     \end{subfigure}
        \caption{\textbf{Left:} Same CMD as in Figure~\ref{panel_2}, plus the cross-match with APOGEE and the PIGS follow-up, colour-coded by their spectroscopic metallicities.
        \textbf{Middle:} Spectroscopic metallicities from APOGEE DR16 (grey) and PIGS (blue) in our Sgr sample. \textbf{Right:} Radial velocities from the same APOGEE and PIGS samples. The stars are well distributed around $140\, \kms$, the mean Sgr radial velocity \citep{1994Natur.370..194I}.}
\label{panel_1}
\end{figure*}

We constrained the photometric variability for the \Gaia photometry using the \Gaia catalogue parameters \texttt{phot\_g\_n\_obs} (the number of observations in the G-band) and \texttt{phot\_g\_mean\_flux\_over\_error} (the mean flux over the error in the G-band), following equations 17 and 18 in \citet{2021MNRAS.508.1509F}. We further clean the sample using the BP$-$RP flux excess: $|C_*|\leq 3\sigma_{C_*}(G)$ \citep{2021A&A...649A...3R}. These two last refinements remove $1.3$ percent of the previous selection. We finally defined the Sgr members with a cut on the proper motions (PMs), as shown in Figure \ref{density}. We select stars within a radius of 0.6 $ \mathrm{mas}\,\mathrm{yr^{-1}}$ of the mean PM of Sgr, corrected for its variation with RA and Dec as reported in the work \citep{2020MNRAS.497.4162V} using the following equations:

\begin{equation}\label{eq:pm}
\begin{split}
    \overline{\mu_{\alpha}}=-2.69+0.009\Delta\alpha-0.002\Delta\delta-0.00002{\Delta\alpha}^{3}\\
    \overline{\mu_{\delta}}=-1.35-0.024\Delta\alpha-0.019\Delta\delta-0.00002{\Delta\alpha}^{3}
\end{split}
\end{equation}

where $\Delta\alpha = \alpha - \alpha_{0}$, $\Delta\delta = \delta - \delta_{0}$ represent the difference in RA and Dec with respect to the Sgr centre. We assume the nuclear globular cluster M54 (NGC6715) to be at the Sgr centre, with coordinates $\alpha_{0} = 283.764^{\circ}$ and $\delta_{0} = -30.480^{\circ}$.

We present the colour-magnitude diagram (CMD) and the Pristine colour-colour diagram for the Sgr selection (after applying the cuts above) in Figure \ref{panel_2}. The Ca H\&K term appears in the y-axis of the colour-colour diagram, making it sensitive to metallicity and creating a spread in the metallicity values along the vertical axis. One can see three distinct blobs in the colour-colour diagram. The largest blob consists of red giant branch stars in Sgr. 

The small blob of stars around $\bprp = 0.8$ and y-axis = 6.5 has values spreading a wide range in parallax compared the rest of the stars. These are likely foreground stars, which we decide to cut using the red dashed line in Figure~\ref{panel_2}, together with all bluer stars which are mostly other types of contamination. With this last cut, our final Sgr sample contains 44\,785 stars. 

The stars in the distinct sequence in the colour-colour diagram between $\sim \bprp$ = 0.9 to 1.2 (between the red and blue dashed lines) do not have a different parallax distribution compared to the main group of stars, and follow the spatial distribution of the main group of stars as well. We hypothesise that these are not foreground stars, nor Sgr red giant branch stars, but rather Sgr helium-burning red clump/horizontal branch stars, which have slightly higher temperatures compared to the normal red giants and hence form a distinct sequence in the colour-colour diagram. The main red clump in Sgr occurs at $G_{0}\sim 17.7$ and it is excluded from our selection, hence the brighter red clump stars must be on the near side of the dwarf galaxy to enter in our  $G_{0}< 17.3$ selection. 

\subsection{Spectroscopic samples} 

We use spectroscopy from two different Sgr samples to evaluate to our Sgr selection. The first sample we use is APOGEE DR16 \citep{2020ApJS..249....3A}, with which we have 568 stars shared with our Sgr selection. Most of these are relatively metal-rich ($\feh > -1.0$), see the middle panel of Figure \ref{panel_1}. We checked the latest APOGEE release as well \citep[DR17,][]{2022ApJS..259...35A}, but did not find any additional stars in common with our Sgr selection.

The second sample we use is a spectroscopic follow-up from PIGS, which contains almost exclusively lower metallicity stars. \citet{arentsen2020pristine} presented the PIGS low- and medium-resolution spectroscopic follow-up of metal-poor inner Galaxy candidates, obtained with AAOmega+2dF on the AAT \citep{2004SPIE.5492..389S, 2002MNRAS.333..279L, 2006SPIE.6269E..0GS}. The low-resolution ($R \sim 1300$) optical and intermediate resolution ($R \sim 11\,000$) calcium triplet spectra were analysed through a full-spectrum fitting using the FERRE code\footnote{FERRE is available at \url{http://github.com/callendeprieto/ferre}} \citep{2006ApJ...636..804A}, providing effective temperatures, surface gravities, metallicities and carbon abundances (see \citealt{arentsen2020pristine} for details). More spectroscopic follow-up was obtained later in 2020, which has been analysed in the same way and has already been used in \citet{arentsen21}. In total, four AAT pointings included dedicated observations of Sgr stars (see their positions in Figure~\ref{fig:footprint}). These have not yet been discussed in any publication. 

The Sgr selection for the follow-up was made by adding two simple \Gaia DR2 cuts to the PIGS selection: $ (\varpi - \epsilon_{\varpi}) < 0.1$ and proper motions within 0.6 $\mathrm{mas\,yr^{-1}}$ of $\mu_{\alpha} = -2.7$ and $\mu_{\delta} = -1.35$ no transformations like Equations~\ref{eq:pm} were applied). We used \Gaia DR2 because the spectroscopic follow-up was done before \Gaia EDR3, as part of the main PIGS follow-up program predating the current work. The Sgr selection was extended to half a magnitude deeper ($G=17.0$) than the main PIGS sample. Whereas the PIGS fields were observed for 2h each, two of the Sgr fields were observed for 3h, one for 2.5h and one for 1.5h. The CMD and metallicities of PIGS stars in our final Sgr selection are shown in the left and middle panels of Figure \ref{panel_1}. The cross-match between the PIGS AAT spectroscopy and our selected Sgr sample results in 426 objects (keeping only stars with good quality spectroscopic parameters, following \citealt{arentsen2020pristine}). Together, the APOGEE and PIGS samples cover the full metallicity range of Sgr.

We present the radial velocities from the spectroscopic PIGS and APOGEE stars in our Sgr sample in the right-hand panel of Figure \ref{panel_1}. For our selection, the median APOGEE and PIGS radial velocities are $\approx 144.5\, \kms$ and $\approx 145.6\, \kms$ respectively. The resulting histogram shows that the radial velocities have a smooth distribution with no clear outliers, around the mean literature Sgr radial velocity of $\sim 140\, \kms$ \citep{1994Natur.370..194I}. The PIGS sample has a slightly higher velocity dispersion than the APOGEE sample, which is in line with the expectation that metal-poor stellar populations are typically older and more pressure-supported. 
We will further discuss this scenario in Section \ref{discussion}.

The colour-coding of CMD shown in the left panel of Figure \ref{panel_1} displays the spectroscopic metallicities of the PIGS and APOGEE stars. The most metal-poor stars are located on the blue side, while the redder part is populated by cooler metal-rich stars. From the metallicity histogram in the middle panel, the dominance of PIGS among the metal-poor stars with respect to the APOGEE sample is clear, the result of PIGS being focused on the search of metal-poor stars in Sgr. A comparison between metallicities from PIGS and APOGEE has been made for bulge stars in \citet{arentsen2020pristine}, finding good agreement -- a dispersion of 0.2~dex, and only a slight systematic offset (with APOGEE being more metal-rich by $0.1-0.2$~dex, depending on the metallicity).

\begin{figure}
     \centering\begin{subfigure}[b]{0.4\textwidth}
         \centering
         \includegraphics[width=\textwidth]{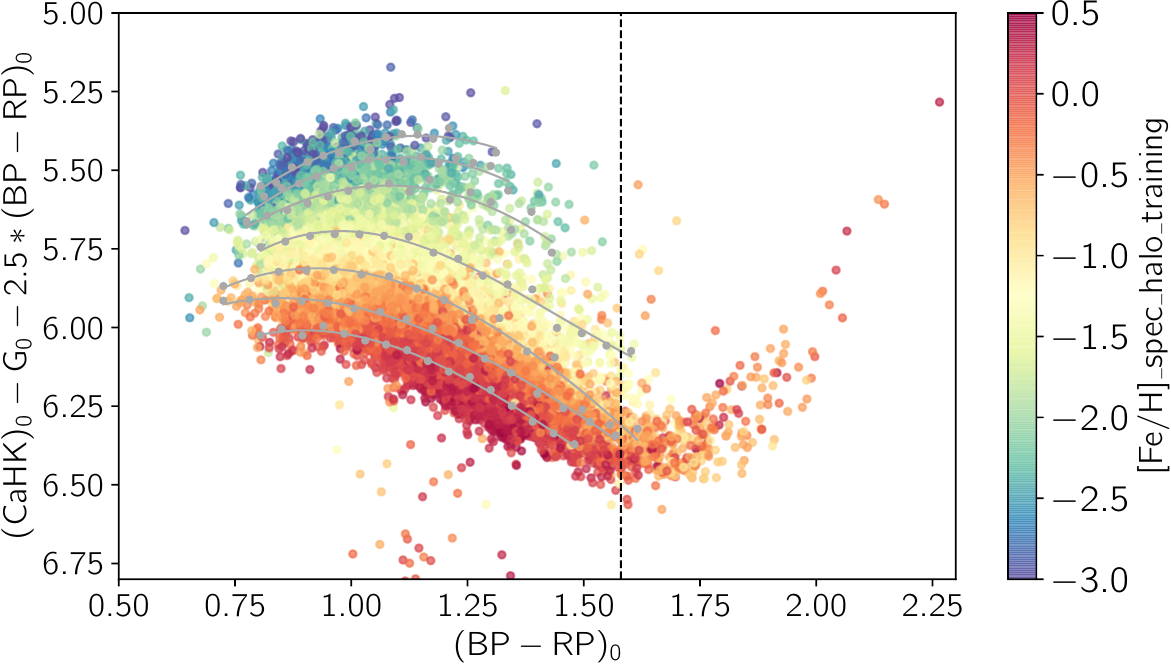}
     \end{subfigure}
     \hfill
     \begin{subfigure}[b]{0.41\textwidth}
         \centering
         \includegraphics[width=\textwidth]{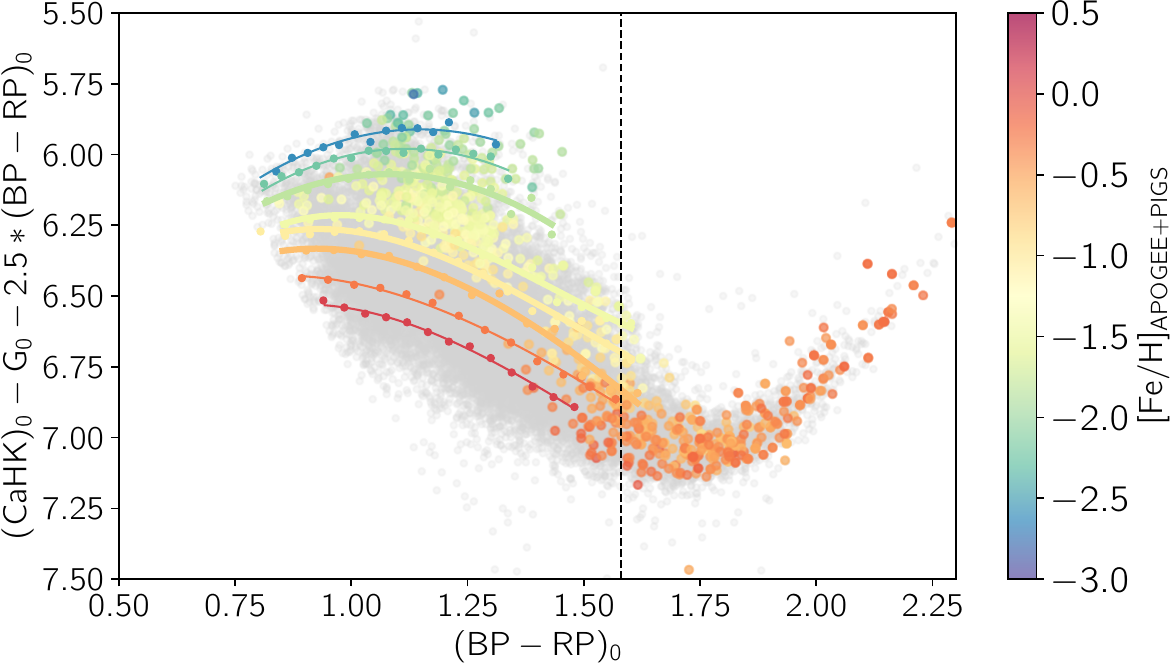}
     \end{subfigure}
     \caption{\textbf{Top:} Colour-colour diagram for the main \Pristine survey spectroscopic training sample, colour-coded by their spectroscopic metallicity. The grey lines are polynomials fitted through the grey points, which are medians of the binned \bprp axis for stars within 0.1~dex of $\feh = -3.0, -2.5, -2.0, -1.5, -1.0, -0.5$ and $0.0$, from top to bottom, respectively. \textbf{Bottom:} Colour-colour diagram for our Sgr sample (full sample shown in grey), with the spectroscopic Sgr sample from APOGEE and PIGS colour-coded by their spectroscopic metallicity. Overplotted are the iso-metallicity lines derived from the training sample, after applying a vertical offset to match with the Sgr spectroscopic metallicities. The thicker lines mark the threshold values used to divide different metallicity populations in this work, namely $\feh = -2.0, -1.5, -1.3$ and $-1.0$. The vertical dashed lines at $\bprp = 1.6$ in both panels limits the area for which we use the iso-metallicity lines (on the left). The stars redder than this will all be classified as metal-rich based on the spectroscopic \feh available.}
\label{CCD_lines}
\end{figure}

\subsection{Spectroscopic calibration sample}

We make use of a training sample with available spectroscopic metallicities present in the footprint of the main \Pristine halo survey \citep{2017MNRAS.471.2587S} to derive iso-metallicity lines for the \Pristine colour-colour diagram, which we will use to calibrate the metallicity scale of our photometric PIGS-Sgr sample. The training sample consists of the main training sample for the \Pristine survey, which has been carefully built to contain many very metal-poor stars (from SEGUE/\citealt{yanny2009segue}, LAMOST/\citealt{li2018lamost} and the dedicated \Pristine follow-up/\citealt{aguado19}), enriched with additional stars from APOGEE DR16 \citep{2020ApJS..249....3A} to extend the training sample to higher metallicities and lower temperatures. Because we only have giant stars in our Sgr sample, we only keep the giants in the training sample ($\logg < 3.8$ and $\teff < 5700$~K). After cross-matching this spectroscopic sample with the most recent internal \Pristine~\CaHK catalogue (already cross-matched with \Gaia), de-reddening it in the same way as our Sgr photometry, and applying the same photometric quality cuts as before, we obtain a sample of $\sim 23000$ giant stars with $-4.0 < \feh < +0.5$. The resulting sample is shown on the \Pristine \CaHK-\Gaia colour-colour diagram in the upper panel of Figure~\ref{CCD_lines}, colour-coded by the spectroscopic metallicities from the training sample. On top of it we show our derived iso-metallicity lines, ranging from $-3.0$ to $0.0$ in steps of $0.5$~dex, which will be further described in the following section.

\section{Metallicity analysis} \label{met_analysis}

The following section reports the metallicity analysis conducted on the photometric Sgr selection with the help of the spectroscopic training sample. With the aim of studying the distribution of different metallicity populations in Sgr, we divide the sample in two main groups and study their spatial distribution. We also fit models to the Sgr stellar density, paying attention to a possible metallicity gradient within Sgr. Finally, we present the spatial distribution of the very metal-poor stars.

\subsection{Derivation of iso-metallicity lines}\label{isolines}

We employ the spectroscopic training sample from the main \Pristine survey to derive iso-metallicity lines in the \CaHK-\Gaia colour colour space, which we will use in our Sgr analysis to divide the sample into various groups of metallicity. Some iso-metallicity lines have been derived for the \Pristine-SDSS colour space before, but not for the Pristine-\Gaia colour-colour space. For this work, we are only interested in giants since only giants are part of our Sgr selection. We binned the colour-colour space in \bprp and selected slices of spectroscopic metallicities within 0.1~dex of a given \feh, with a minimum number of 5 stars per \bprp bin. To derive an iso-metallicity line we determined the median y-axis value in each bin and successively fit a 2nd order polynomial to these points for $\feh \leq -1.8$ and a 3rd order for $\feh \geq -1.7$. The resulting iso-metallicity lines are show in the upper panel of Figure~\ref{CCD_lines}, on top of the spectroscopic training sample from the main \Pristine survey. The lines range from $\feh = -3.0 $ to $\feh = 0.0$ dex, in steps of 0.5 dex. We limit the polynomials to $\bprp < 1.6$, since for redder colours the metal-poor and metal-rich stars start overlapping and crossing. 

The derived iso-metallicity lines are on the main survey calibration scale for the \CaHK photometry, but the Sgr-PIGS photometry is on a different scale, offset by a constant. To determine the offset we used the available Sgr spectroscopy from APOGEE and the PIGS/AAT data, which is shown and colour-coded by metallicity in the bottom panel of Figure~\ref{CCD_lines}. We compute the offset between a given iso-metallicity line and the stars in the Sgr spectroscopic sample falling in the same metallicity range. We used Sgr stars with $-1.6 < \feh < -1.4$, $-1.9 < \feh < -1.7$, and $-2.1 < \feh < -1.9$, and computed the difference between these stars and the iso-metallicity lines at $\feh = -1.5, -1.8$ and $-2.0$, respectively, finding an average shift of 0.52 mag. In the bottom panel of Figure~\ref{CCD_lines}, the same iso-metallicity lines from the upper panel (plus one at $\feh = -1.3 $ represented by the thicker yellow line) are shown, now shifted with the offset derived above to match the Sgr colour-colour space. They are colour-coded by the corresponding spectroscopic metallicity. 

\begin{figure*}
     \begin{subfigure}[b]{0.27\textwidth}
         \centering
         \includegraphics[width=\textwidth]{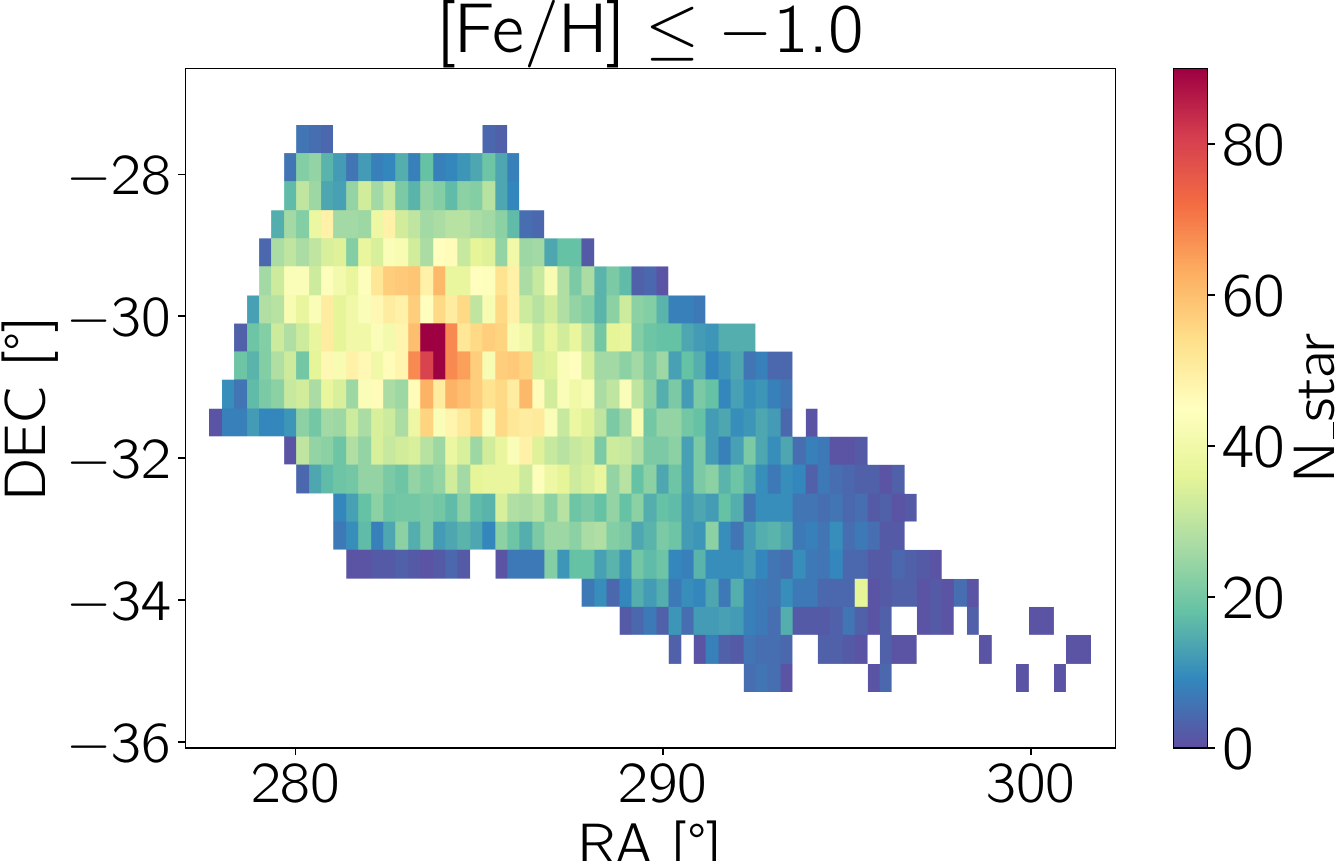}
     \end{subfigure}
     \hspace{0.05cm}
     \vspace{0.1cm}
     \begin{subfigure}[b]{0.27\textwidth}
         \centering
         \includegraphics[width=\textwidth]{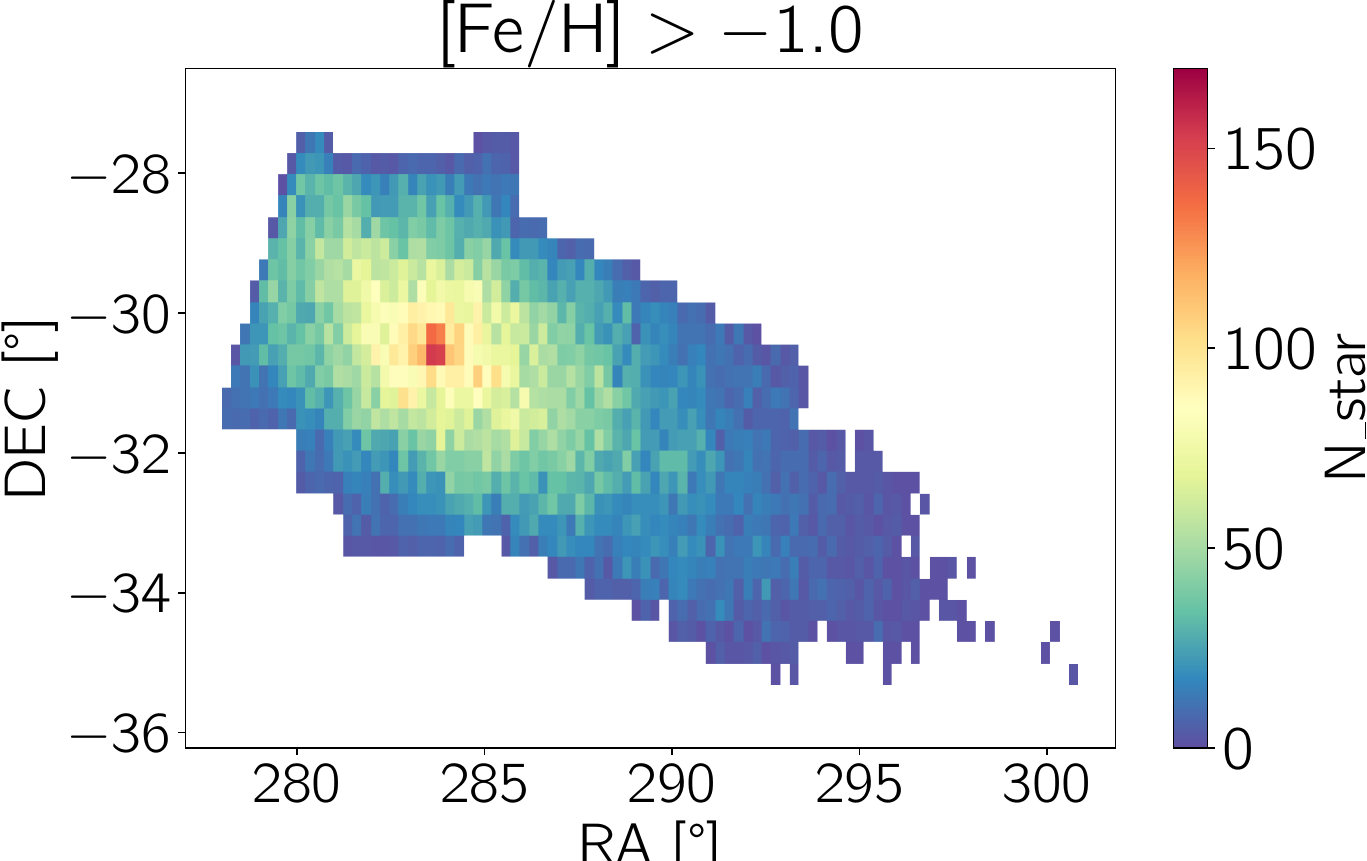}
     \end{subfigure}
     \hspace{0.05cm}
     \vspace{0.1cm}
      \begin{subfigure}[b]{0.44\textwidth}
         \centering
         \includegraphics[width=\textwidth]{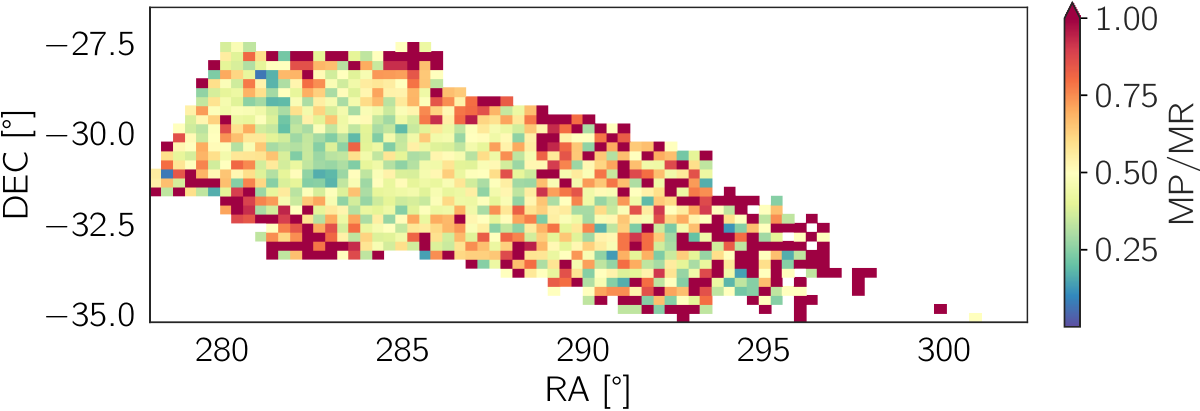}
     \end{subfigure}
     \hspace{0.05cm}
     \vspace{0.1cm}
     \begin{subfigure}[b]{0.27\textwidth}
         \centering
         \includegraphics[width=\textwidth]{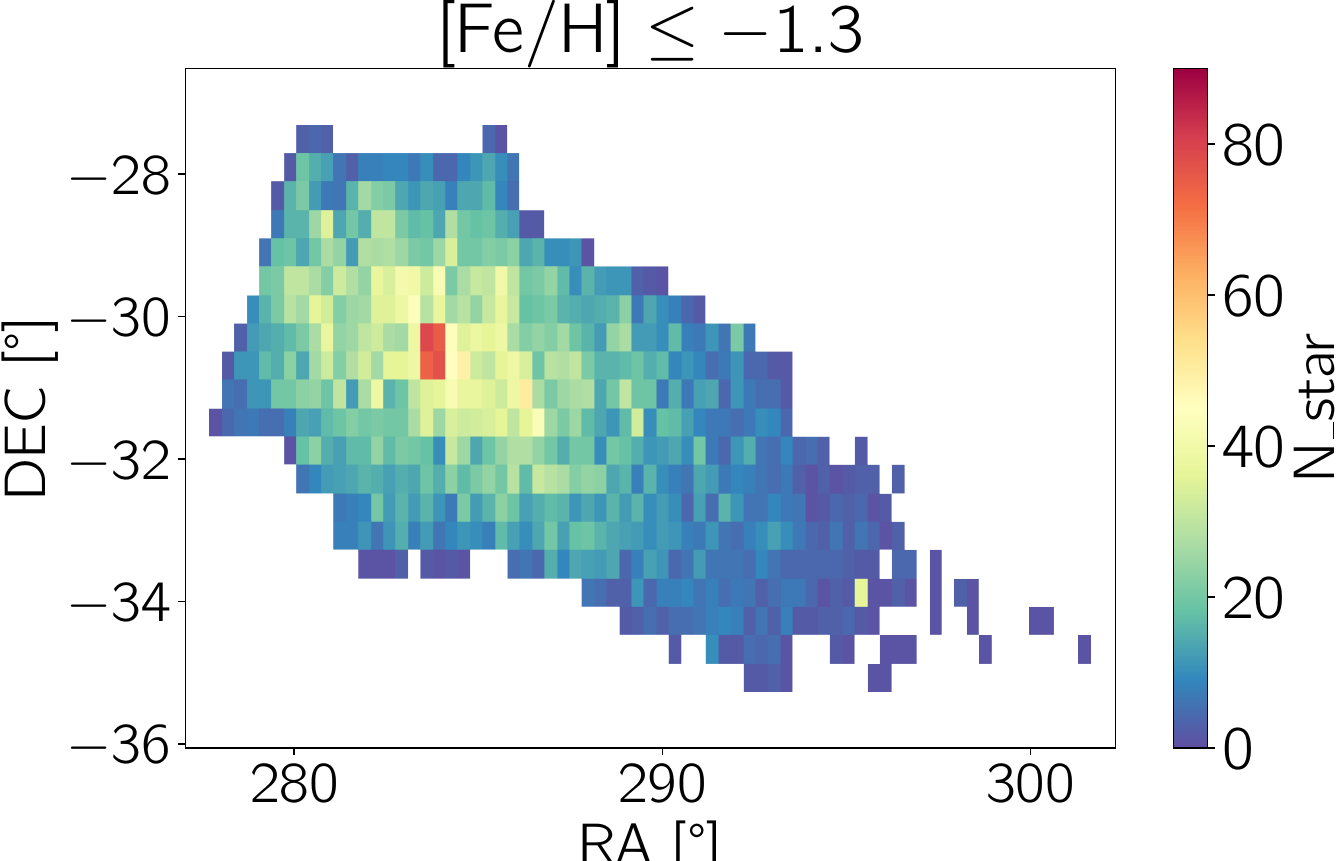}
     \end{subfigure}
     \hspace{0.05cm}
     \vspace{0.1cm}
     \begin{subfigure}[b]{0.27\textwidth}
         \centering
         \includegraphics[width=\textwidth]{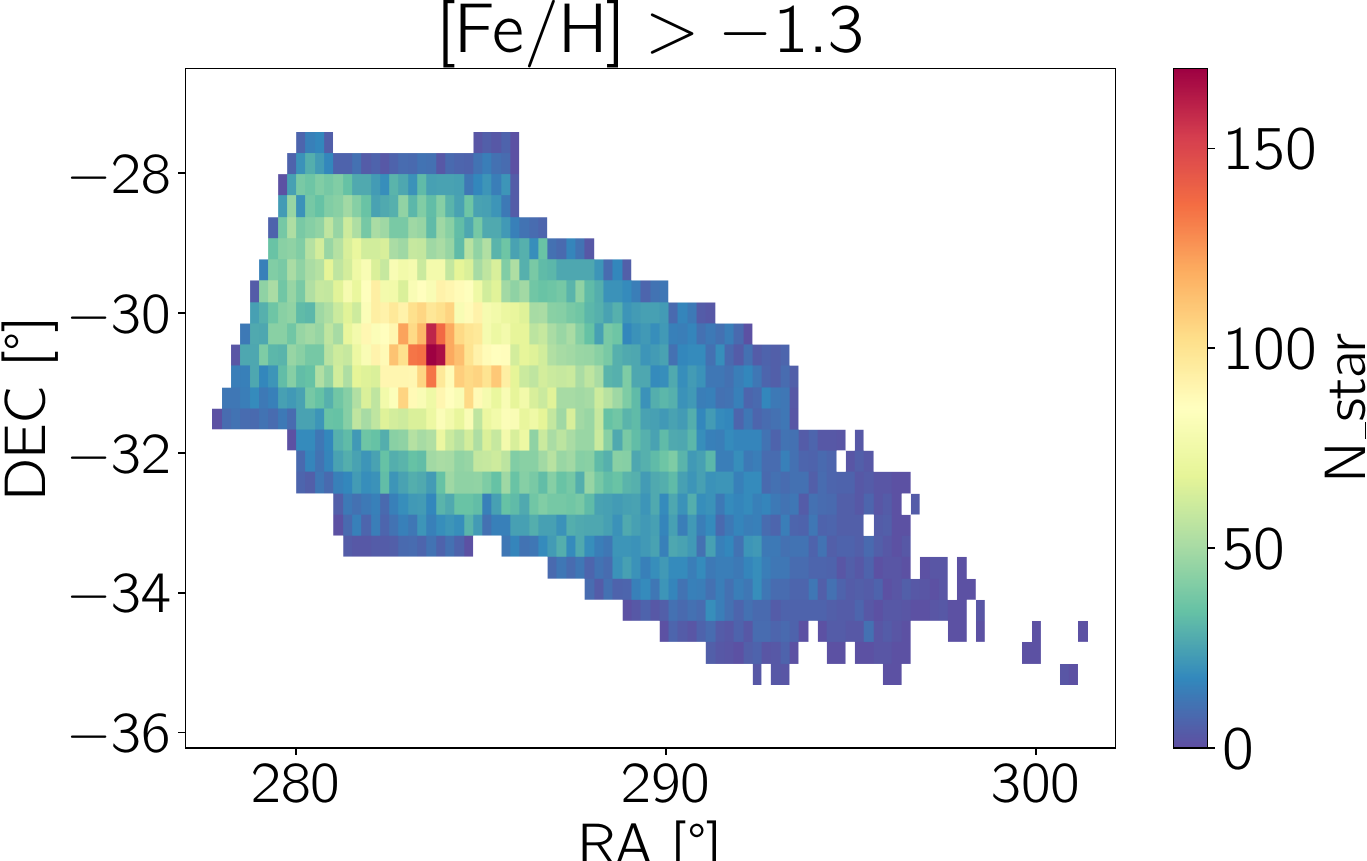}
     \end{subfigure}
     \hspace{0.05cm}
     \vspace{0.1cm}
      \begin{subfigure}[b]{0.44\textwidth}
         \centering
         \includegraphics[width=\textwidth]{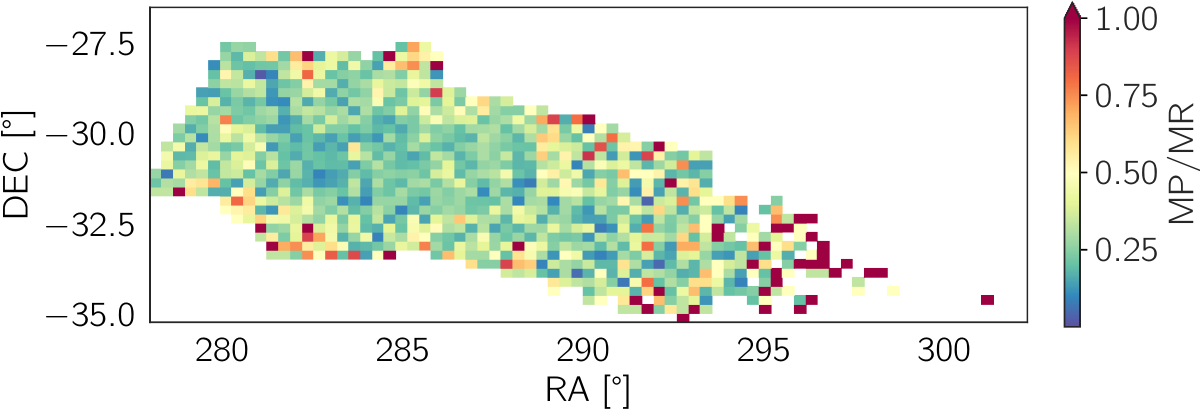}
     \end{subfigure}
     \hspace{0.05cm}
     \vspace{0.1cm}
     \begin{subfigure}[b]{0.27\textwidth}
         \centering
         \includegraphics[width=\textwidth]{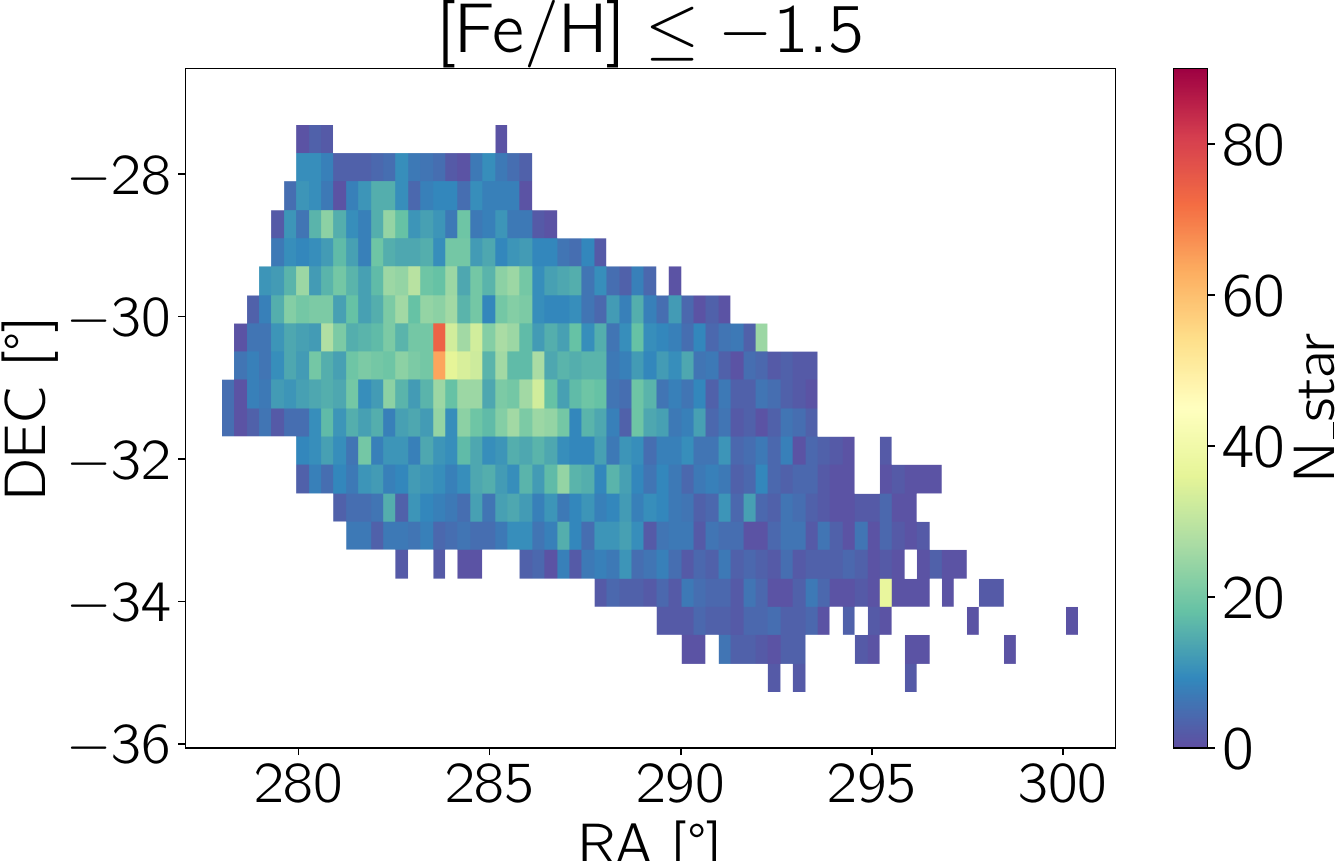}
     \end{subfigure}
     \hspace{0.05cm}
     \vspace{0.1cm}
     \begin{subfigure}[b]{0.27\textwidth}
         \centering
         \includegraphics[width=\textwidth]{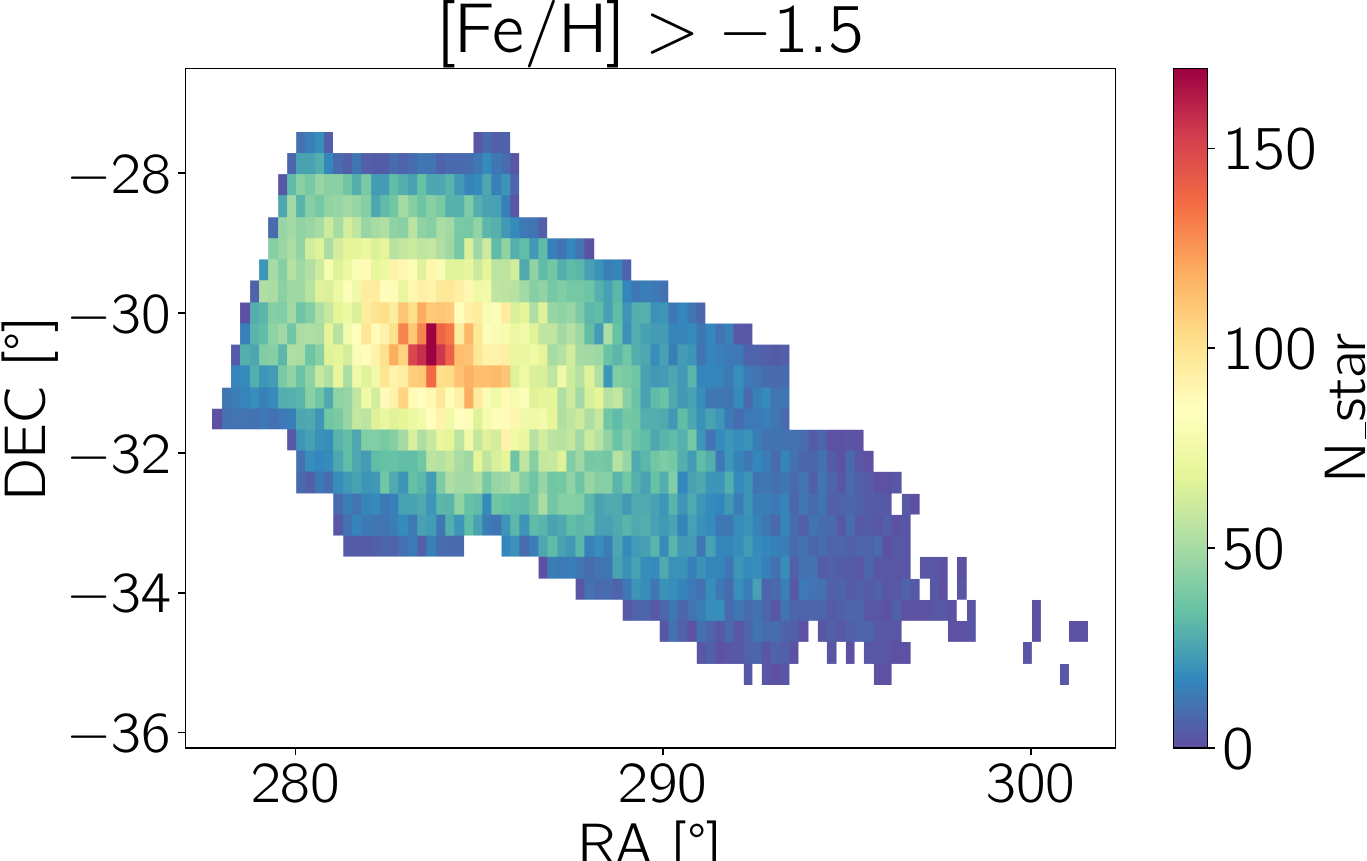}
     \end{subfigure}
     \hspace{0.05cm}
     \vspace{0.1cm}
      \begin{subfigure}[b]{0.44\textwidth}
         \centering
         \includegraphics[width=\textwidth]{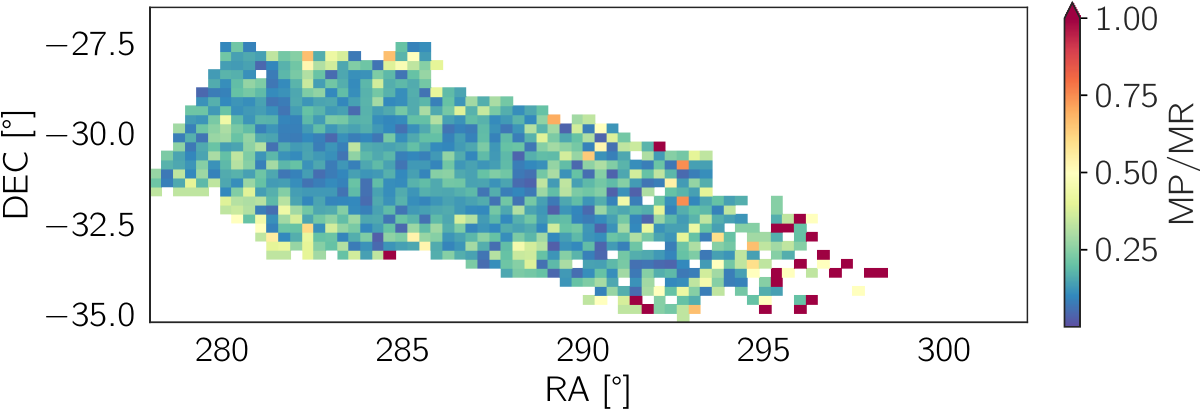}
     \end{subfigure}
     \caption{The first column represents the density distribution for the selected MP stars in equatorial coordinates. At the top the MP group is derived using the $\feh = -1.0$ line, the middle plot is determined with the -1.3 metallicity line, while the selection at the bottom with $\feh = -1.5$. Similarly, the density maps for the various MR-selected populations ($\feh >-1.0; \feh > -1.3; \feh >-1.5$) are shown in the middle column. The right column displays the 2D histograms of the ratios between MP and MR stars (same selections adopted for the density maps) in RA and DEC. Each bin of $0.33^{\circ}$ is colour-coded by the ratio value. To have a clear spread in the ratio range, the $\mathrm{MP/MR} > 1$ are set to 1.}
\label{MPMR}
\end{figure*}

We found that the shift between the Sgr spectroscopic metallicities and the iso-metallicity lines derived from the training sample depends on the exact metallicity range that is used. We hypothesize that this is connected to a significant difference in the [$\alpha$/Fe] abundances between MW stars and Sgr stars. In the training sample from the main \Pristine survey, the reddest part of the colour-colour diagram (with $ \bprp \gtrsim 1.5$) splits into two sequences for giants when colour-coded by [$\alpha$/Fe] from APOGEE. Their alpha abundances are representative of thin and thick disc stars. The Sgr [$\alpha$/Fe] is significantly lower than both of those \citep{2017ApJ...845..162H}. Some discussion on this can be found in the Appendix. For this reason, we will not be determining individual photometric metallicities for each star in this paper, because it is not clear exactly what scale they would be on. The choice of cuts performed employing these iso-metallicity lines is extensively discussed in further sections, and we show that our main interpretations do not depend on the details.

\subsection{Metallicity division for Sgr stellar populations}\label{division}

The iso-metallicity lines offer an effective way to separate the Sgr sample into two main populations, one metal-rich (MR) and one metal-poor (MP) population. Looking at how the choice of different iso-metallicity lines has an impact on the metallicity distribution within the Sgr core gives an insight about the role of the selection selection effect on the final results.  

First, we use the iso-metallicity line at $\feh = -1.0$ dex (thicker orange line in Figure \ref{CCD_lines}). With this division, 14670 stars compose the $\feh < -1.0$ group and 30115 stars are identified with $\feh > -1.0$. Next, we test the separation using $\feh = -1.3$ (the line is coloured in yellow in Figure~\ref{CCD_lines}), resulting in a metal-poor population of 9719 stars and a metal-rich population of 35066 stars. Finally, we also tested a division at $\feh = -1.5$ (thicker light-green line), obtaining 6195 stars with $\feh < -1.5$ and 38590 stars with $\feh > -1.5$. The polynomial division is valid until $\bprp = 1.6$. 
Thanks to the APOGEE dataset, it is possible to classify the stars sitting on the redder part of the diagram (on the right of the dashed line in Figure \ref{CCD_lines}) as all being part of the population with higher metallicity, as there are 305 spectroscopic stars with $\bprp > 1.6$ with $\feh>-1.0$. We include the reddest stars in our metal-rich populations in the remainder of this work.

\begin{figure}
     \begin{subfigure}[b]{0.50\textwidth}
         \includegraphics[width=\textwidth]{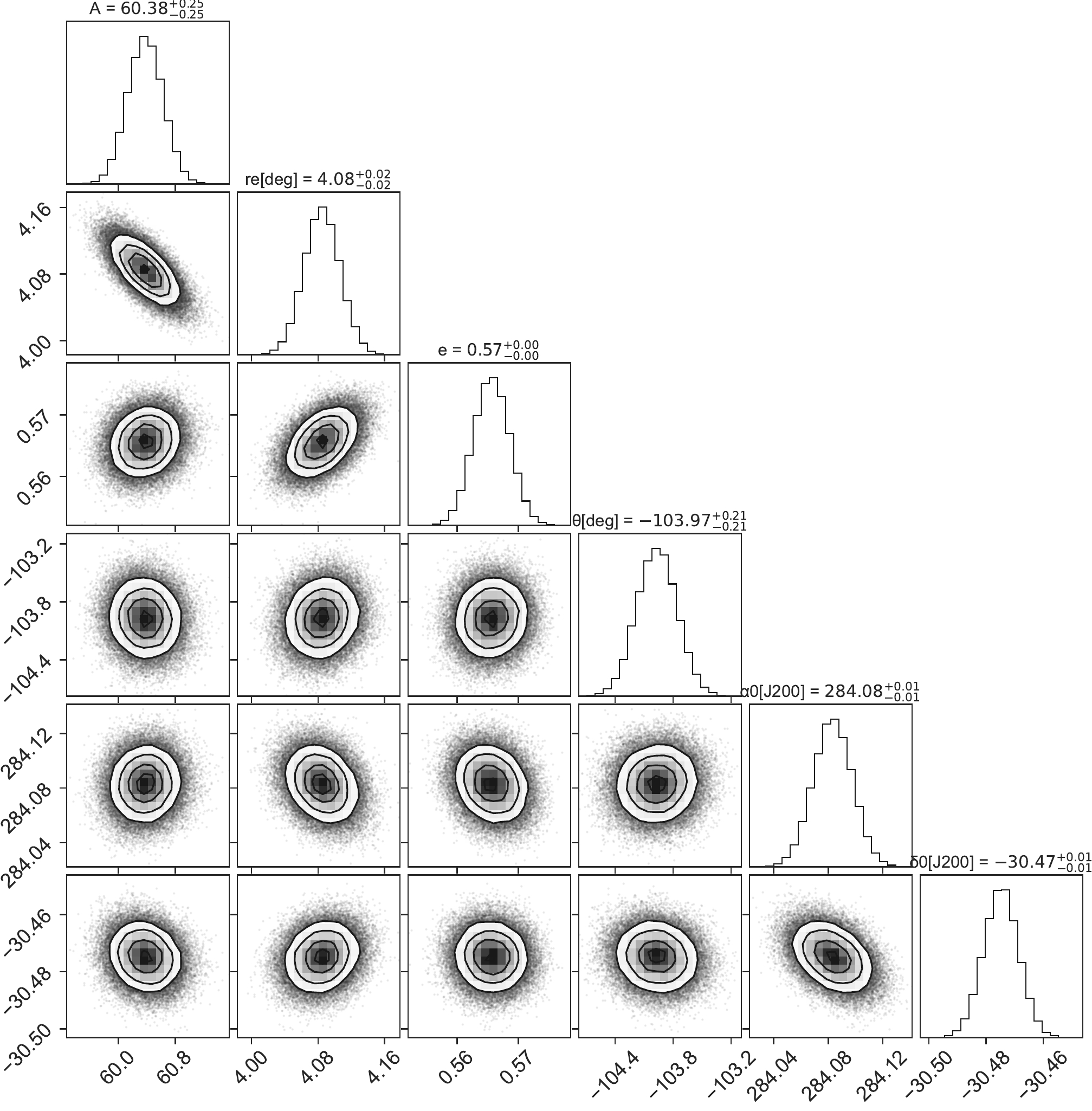}
     \end{subfigure}
     \hspace{0.15cm}
     \begin{subfigure}[b]{0.50\textwidth}
         \includegraphics[width=\textwidth]{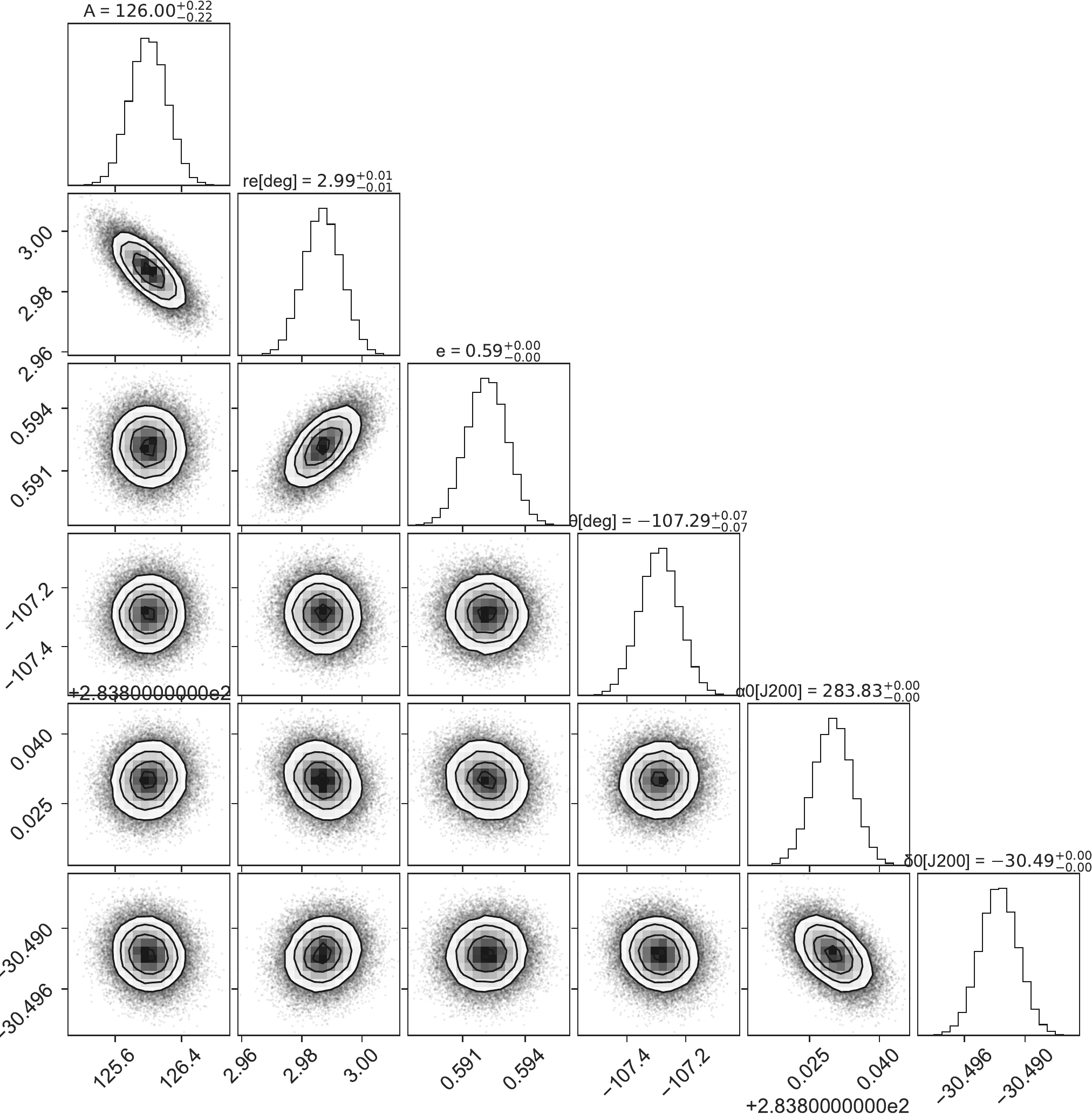}
     \end{subfigure}
        \caption{\textbf{Top:} Corner plot illustrating the results of the our model fit for the MP population ($\feh < -1.3$). They show the covariance between the set of parameters, hinting a correlation between \textit{re} and \textit{A}, \textit{re} and \textit{e}, and between the coordinates of the centre of the distribution ($\mathrm{ra_{0}}\, , \mathrm{dec_{0}}$). \textbf{Bottom:} Same, but for the Sgr MR population ($\feh > -1.0$).}
\label{models1}
\end{figure}

\begin{figure*}
     \begin{subfigure}[b]{0.88\textwidth}
        % \centering
         \includegraphics[width=\textwidth]{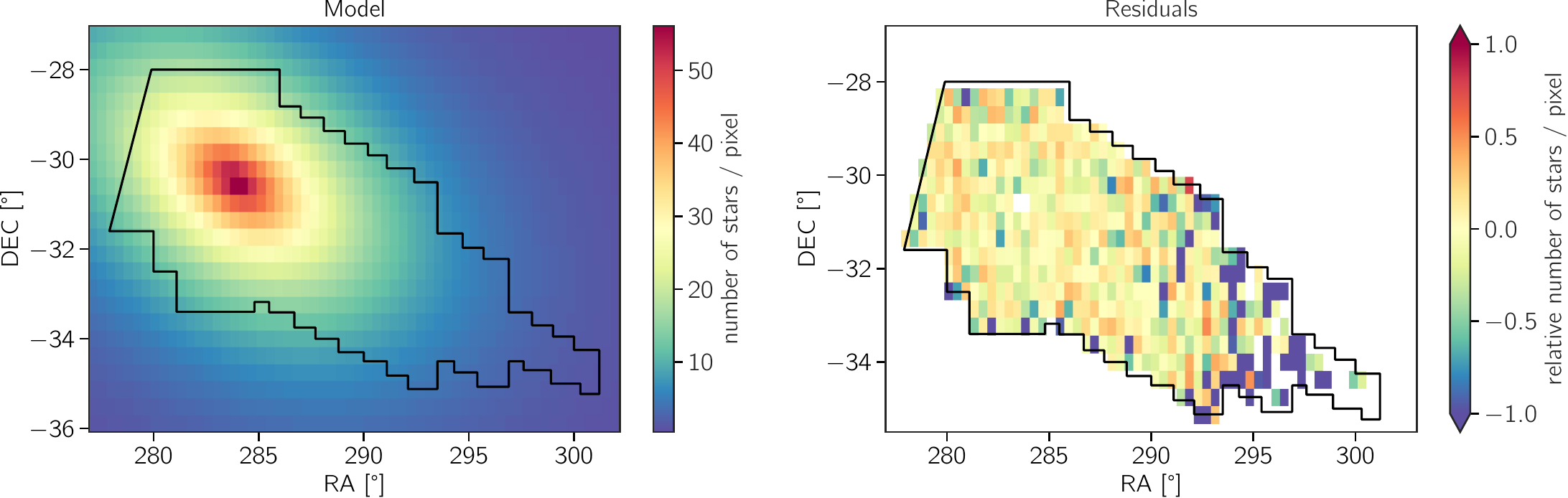}
     \end{subfigure}
     \hspace{0.15cm}
     \begin{subfigure}[b]{0.88\textwidth}
         %\centering
         \includegraphics[width=\textwidth]{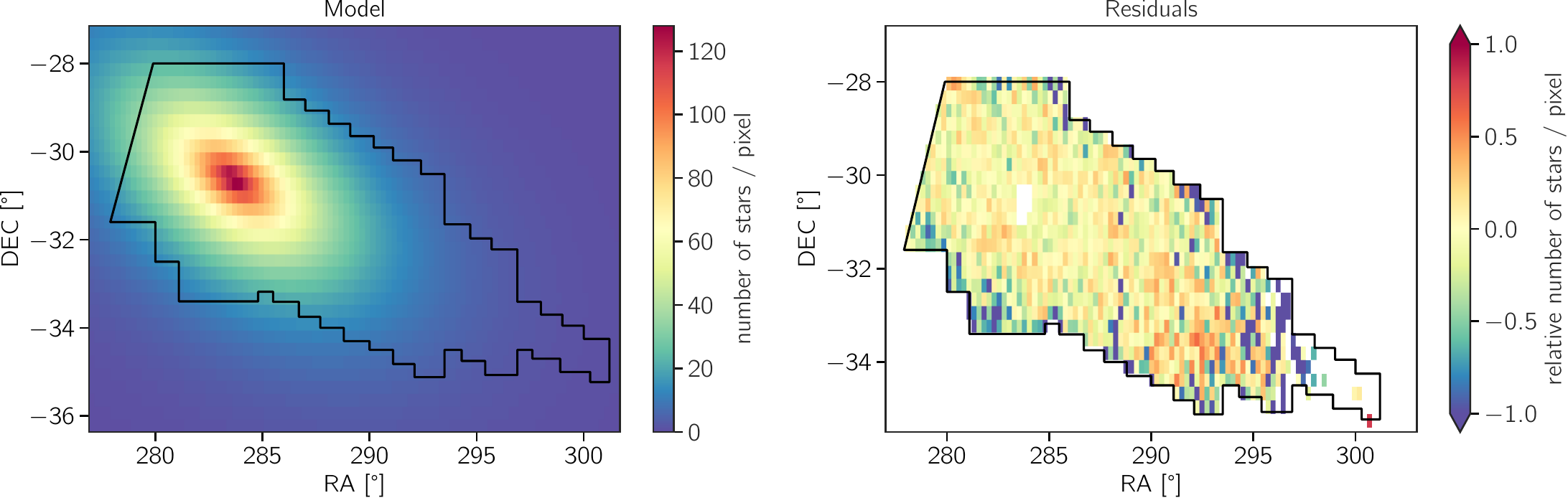}
    \end{subfigure}
        \caption{\textbf{Left:} Our best models for the MP population ($\feh < -1.3$, top) and MR population ($\feh > -1.0$, bottom), colour-coded by the number of stars. The PIGS Sgr footprint indicated by a black solid line. \textbf{Right:} Residuals of the fits in relative scale (difference between the data and the model divided by the number of stars in each pixel). The colour bar is restricted between $-1$ and $+1$ for a clearer visualisation of their distribution along the footprint. The corresponding observed data are shown in the panels of Figure~\ref{MPMR}.} 
\label{models}
\end{figure*}

\subsection{Spatial distributions} \label{pattern}

Next, we build density maps of the metal-poor and metal-rich populations defined in Section \ref{division} to study their spatial distributions. The spatial distributions are shown in the left and middle columns in Figure~\ref{MPMR} for the three different metallicity separations, all of which practically show the same pattern: a higher stellar density is found in the centre of Sgr around M54, while the number of observed stars drops moving outwards from the centre to the onset of the stream. One thing to notice is that the more metal-rich component appears to be more centrally concentrated compared to the more spread-out metal-poor population. The very central density of the two metallicity populations is dominated by the central globular cluster M54 \citep[with average $\feh \simeq -1.55$,][]{2010ApJ...714L...7C}. This relates to the very complicated metallicity distribution function of the central region of Sgr influenced by the presence of the nuclear star cluster \citep{2008AJ....136.1147B,2020ApJ...892...20A}.

We use Figure~\ref{MPMR} to investigate the presence of a metallicity gradient. The ratios between the metal-rich and metal-poor populations are shown in the right-hand column. The three rows corresponds to different selections obtained using iso-metallicity lines fitted for various metallicity values, i.e. $\feh = -1.5, -1.3\, \mathrm{and} -1.0$. The histogram shows that the relative fraction of metal-poor objects is higher further away from the centre, while the central area presents on average a lower fraction. This effect is less visible, although always present, as the metallicity limit for dividing the populations get lower. Indeed for the division at $\feh = -1.5$ the number of stars with $\feh > -1.5$ severely outnumber the metal-poor counterpart and the values of their ratios are shifted towards much lower values. 

For the analysis that follows in this work, we chose to use the stars with $\feh < -1.3$ as the metal-poor (MP) populations, while for the metal-rich (MR) group we set $\feh > -1.0$. In this way we ensure that the division between the two groups is more clean and not dominated by stars that are close to the dividing line. We check the photometric selection with the help of the APOGEE and PIGS spectroscopy available, which covers most of the investigated metallicity range. We find that $\sim 98\%$ of spectroscopic stars with $\feh<-1.3$ are part of our MP population, while $\sim95\%$ of the stars with spectroscopic metallicities $> -1.0$ are part of our MR population. We have tested that this characteristic does not depend strongly on the metallicity limit used to separate the MP and MR components. The main result, that the MP/MR ratio increases further from the Sgr centre, is visible also when using even lower $\feh $ values for defining the MP population. 

Fixing the limit for the MP population at $\feh = -1.3$ can also be justified by the estimated location of the alpha-knee in Sgr, although its position is still under debate  as well as the corresponding \feh value. It has been derived for M54 to occur at $\feh \sim -1.3 / -1.4$ \citep{2010A&A...520A..95C} and at $\feh \sim -1.3 $ for the Sgr stellar streams \citep{2014MNRAS.443..658D}. The alpha knee is where the turn-over happens in the $\feh - \alphafe$ diagram, and is linked to the star formation rate of a system. Before this change in the \alphafe abundance, the chemical enrichment is mostly subjected to the presence of the supernovae type II (SNII) from massive stars, which have much shorter timescales than supernovae type Ia (SNIa) with lower mass progenitors. A galaxy that builds up and maintains metals and gas before SNIa start to contribute to the gas enrichment will reach a higher metallicity of the knee, compared to a galaxy which looses metals due to galactic winds, or does not show an efficient SFH \citep{2020ARA&A..58..205H}. 

\subsection{Quantifying the differences between MP and MR populations} \label{model_sec}

\subsubsection{Method}

To quantify the structural differences between the MP and MR populations, we fit a model to their spatial distributions. 
We bin the Sgr footprint in pixels of a few tens of arc-minutes, $\sim 20'$ for the MP and $\sim 15'$ MR, as the stellar density of this latter population is higher than the one of the metal-poor component. We follow the approach of \citet{2008ApJ...684.1075M} and express the Sgr stellar density as:
\begin{equation}\label{eq1}
    % \mathrm{N_{model}(p_{1},...,p_{j})} =
    \mathrm{N_{i}} = 
    \mathrm{A_{0}}\exp^{-\frac{\mathrm{r_{i}}}{\mathrm{r_{e}}}}%+\Sigma_{\mathrm{m}}
\end{equation}
where i indicates the pixel, $\mathrm{A_{0}}$ is the central density, $\mathrm{r_{e}}$ is the exponential scale radius, and $\mathrm{r_{i}}$ is the elliptical distance of each pixel with respect to the centre of the distribution. The r depends on the ellipticity e and the position angle $\theta$ by:
\begin{equation}
    \mathrm{r_{i}} =\left( \frac{1}{1-\mathrm{e}}(\mathrm{x_{i}}\cos{\theta}-\mathrm{y_{i}}\sin{\theta})^{2}+(\mathrm{x_{i}}\sin{\theta}+\mathrm{y_{i}}\cos{\theta})^{2}\right)^{1/2}
\end{equation}
Where x$_i$ and y$_i$ are related to the right ascension ($\alpha$) and declination ($\delta$) and to the tangential plane of the sky by:
\begin{equation}\label{eq:xy}
\begin{cases}
\mathrm{x_{i}} =\frac{\cos(\alpha_{i})\sin(\alpha-\alpha_{0})}{\cos(\delta_{0})\cos(\delta)\cos(\alpha-\alpha_{0})+\sin(\delta)\sin(\delta_{0})}\\[0.2cm]
\mathrm{y_{i}} = \frac{\sin(\delta_{0})\cos(\delta)\cos(\alpha-\alpha_{0})-\cos(\delta_{0})\sin(\delta)}{\cos(\delta_{0})\cos(\delta)\cos(\alpha-\alpha_{0})+\sin(\delta)\sin(\delta_{0})}
\\
\end{cases}
\end{equation}\label{projected}
with the centre at ($\alpha_{0}$, $\delta_{0}$).

The best set of parameters were obtained using Markov Chain Monte Carlo sampling, using the \texttt{emcee} \footnote{\url{https://emcee.readthedocs.io/en/stable/}} package. We fit a model for the stellar density of the metal-poor and metal-rich populations, with the iso-metallicity lines at $\feh = -1.3$ and $\feh = -1.0$ as discriminant. For each fit, we used 64 walkers and 30000 steps. We excluded all pixels more than 50\% outside of the PIGS-Sgr footprint (the black solid line in Figure~\ref{models}), the density of pixels with 50\%--100\% within the footprint has been scaled according to the fraction inside the footprint. To not bias the fit due to the globular clusters in the footprint, we avoided pixels where M54 is located using a radius of $0.4^{\circ}$ around the coordinates of the centre \citep[283.762$^\circ$,-30.479$^\circ$ ][]{1996AJ....112.1487H}. For the metal-poor population we also removed the contribution of the two globular clusters Arp2 and Ter8 \citep{2020A&A...636A.107B}, fixing the 2 centres at ($292.183^{\circ}$ ; $-30.356^{\circ}$) and ($295.433^{\circ}$;$-33.999^{\circ}$) and using a radius of $0.2^{\circ}$ and $0.3^{\circ}$ for Arp2 for Ter8, respectively. 

\begin{table}
\centering
\begin{tabular}{ |c|c|c| } 
\hline\\[-3ex] 
\multicolumn{3}{|c|}{Metal-poor ($\feh < -1.3$)}\\[-1ex] 
\hline
\hline
 $\alpha_{0}$(J2000) & $\delta_{0}$ (J2000) & $A_{0}$ \\[0.04cm]
 $284.083^{+0.014}_{-0.014}$  & $-30.475^{+0.006}_{-0.006}$  & $60
 .375^{+0.250}_{-0.247}$ \\[0.025cm] 
 $r_{e}$ (deg.) & e & $\theta$ (deg.)  \\ [0.025cm]
 $4.085^{+0.022}_{-0.021}$ &$0.566^{+0.003}_{-0.003}$ &$ -103.971^{+0.208}_{-0.206}$ \\
\hline\\[-3ex] 
\multicolumn{3}{|c|}{Metal-rich ($\feh > -1.0$)}\\[-1ex] 
\hline
\hline
 $\alpha_{0}$(J2000) & $\delta_{0}$ (J2000) & $A_{0}$ \\[0.04cm]
 $283.830^{+0.004}_{-0.004}$  & $-30.493^{+0.002}_{-0.002}$  & $126.008^{+0.222}_{-0.225}$ \\[0.025cm] 
 $r_{e}$ (deg.) & e & $\theta$ (deg.)  \\ [0.025cm]
 $2.987^{+0.006}_{-0.006}$ &$0.592^{+0.001}_{-0.001}$ &$ -107.289^{+0.067}_{-0.068}$ \\
 \hline\\[-3ex]
 \multicolumn{3}{|c|}{$\feh < -1.5$}\\[-1ex] \hline
\hline
 $\alpha_{0}$(J2000) & $\delta_{0}$ (J2000) & $A_{0}$ \\[0.04cm]
 $284.060^{+0.024}_{-0.024}$  & $-30.483^{+0.011}_{-0.010}$  & $30.601^{+0.226}_{-0.225}$ \\[0.025cm] 
 $r_{e}$ (deg.) & e & $\theta$ (deg.)  \\ [0.025cm]
 $3.928^{+0.037}_{-0.036}$ &$0.566^{+0.005}_{-0.005}$ &$ -103.955^{+0.359}_{-0.361}$ \\
 \hline\\[-3ex]
 \multicolumn{3}{|c|}{Very metal-poor $\feh < -2.0$}\\[-1ex] \hline
\hline
 $\alpha_{0}$(J2000) & $\delta_{0}$ (J2000) & $A_{0}$ \\[0.04cm]
 $283.815^{+0.111}_{-0.111}$  & $-30.396^{+0.058}_{-0.053}$  & $9.722^{+0.306}_{-0.300}$ \\[0.025cm] 
 $r_{e}$ (deg.) & e & $\theta$ (deg.)  \\ [0.025cm]
 $4.400^{+0.185}_{-0.173}$ &$0.551^{+0.022}_{-0.023}$ &$ -104.506^{+1.704}_{-1.762}$ \\
 \hline
\end{tabular}
\caption{\label{tab:table-name} Structural parameters for different metallicity populations derived through our $\chi^{2}$ minimisation.}
\end{table}

We tested the inclusion of a Galactic background in the model, using an exponential function dependent on the Galactic latitude,
but we found it to be unconstrained in the fitting procedure. This hints at a very efficient cleaning of MW contamination in our member selection, which we further discuss in Section~\ref{discussion}.

\subsubsection{Results}

The resulting parameters of our fits are summarised in Table~\ref{tab:table-name} and Figure~\ref{models1}, which gives the marginalised two dimensional posterior distributions of the set of parameters. Some correlation can be seen between $\alpha_{0}$ vs. $\delta_{0}$, \textit{re} vs. \textit{A}, and \textit{re} vs. \textit{e}, but the fitted parameters are overall well-constrained. Table~\ref{tab:table-name} also reports the values for $\feh < -1.5$ and $< -2.0$. For this last category the pixel sizes is increased to $\sim 30'$ and the lower number of stars is reflected in the uncertainties. 

Fewer stars are included in the lower metallicity categories resulting in the decrease of the central density (\textit{A}). We find that the scale radius ($r_{e}$) shifts significantly towards greater values for the lower metallicity populations. All other model parameters only show minor changes, which are likely not significant. We suspect that the uncertainties on the model parameters are underestimated, which we will briefly get back to in the discussion. The change of the position angle of $\approx 3.3 $ degrees between the MR and MP populations represents a change of only a small fraction, and similarly for the eccentricities. 

The most visible difference between the centres of the main MP and MR populations concerns the RA component: $0.253^{0.028}_{-0.028}$ degree, but this is still only a small change that is unlikely to be significant. Our derived centres differ from the coordinates of M54 at maximum $\sim 0.31^\circ$ in RA for the MP stars, and $\sim 0.07^\circ$ for the MR population. \cite{2021ApJ...908..244D} derived the central coordinates of Sgr (excluding M54) to be $(\alpha, \delta) = (283.945^\circ, -30.647^\circ)$, which differs from our centres by $(\Delta \alpha, \Delta \delta) = (0.138^\circ, 0.172^\circ)$ for the MP and $(\Delta \alpha, \Delta \delta) = (0.115^\circ, 0.154^\circ)$ for the MR populations.

The spatial distributions of the models and residuals are shown in Figure \ref{models}. The residuals show structure for both the MP and MR fits, especially for the regions with RA $> 290^{\circ}$. 
Our elliptical model appears to be too simplistic to describe the distribution of stars in Sgr, which is not surprising given the complex, disrupting nature of the system. We do not expect the residuals to be related to the \Gaia scanning law. The effect of the scanning law is strongest for faint stars (closer to the magnitude limit in \Gaia, $G \gtrsim 20$), producing inhomogeneities on the scale of $\sim$1 degree \citep{2021A&A...649A...5F}. In this work, we only use relatively bright stars ($G_0 < 17.3$) and do not make any strong cuts on any of the \Gaia uncertainties.

\subsection{Very metal-poor stars}

The study of iron-depleted stars is of great interest for reconstructing the history of the Sgr galaxy, as they carry essential information to draw the story line of the early evolution of their host galaxy. Different investigations of VMP stars in Sgr over the years have led to an improved understanding of the early evolution of this dissolving system \citep{2008AJ....136.1147B,2018ApJ...855...83H, 2019ApJ...875..112C, 2020ApJ...901..164C}, but these are based on just a handful of known very metal-poor stars. The blue, metal-poor region of the Sgr CMD overlaps with the giant branch of the Galactic bulge, and before the \Gaia data became available it was difficult to disentangle these. \citet{2019ApJ...875..112C} and \citet{ 2020ApJ...901..164C} used the \Gaia data in combination with narrow-band SkyMapper $v$ photometry to identify 22 metal-poor stars ($-3.10 < \feh < -1.45$) in Sgr, which was successful but still resulted in only a small sample. 

The spectroscopic Sgr-PIGS follow-up sample contains 100 stars with $\feh_\mathrm{spec} < -2.0$, this is the largest spectroscopic VMP Sgr sample to date. Our full photometric Sgr-PIGS data is an excellent source for more VMP stars. We selected VMP stars following the same approach explained in Section \ref{division}, but this time using the iso-metallicity line at $\feh = -2.0$ (see blue line in Figure \ref{CCD_lines}). Our VMP sample consists of 1150 stars, which is the largest sample of VMP candidates with $\feh < -2.0$ in Sgr to date. 

\begin{figure}
\includegraphics[width=0.45\textwidth]{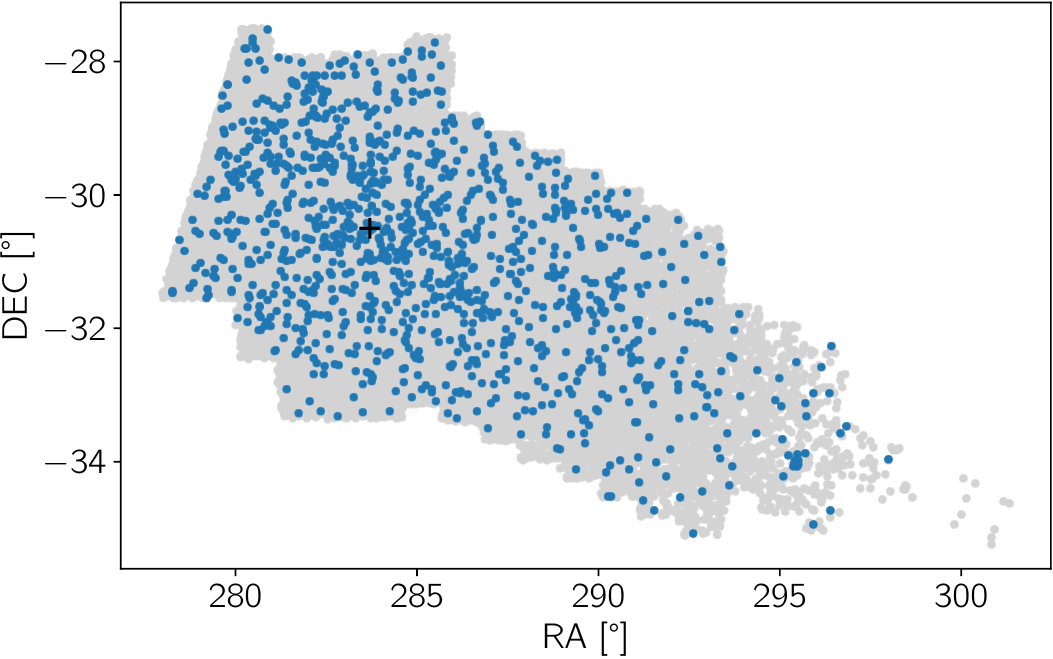}
\caption{Our selected sample of 1150 VMP stars (in blue) on top of the full Sgr sample shown in grey. The VMP stars are extended relatively uniformly over the dwarf galaxy, with a higher density around the centre of Sgr (M54 is indicated with a black cross). }
\label{VMP} 
\end{figure}

The distribution of the VMP stars is shown in Figure \ref{VMP}. Almost the entire selection ($>99.9\%$) is far beyond the tidal radius of M54 (7.5'), therefore not associated to the globular cluster \citep{2010arXiv1012.3224H}. It is clear that these ancient stars are located at all radii but do follow the overall Sgr density distribution, the stellar density being higher in the centre. Around RA $\sim 295^{\circ}$, an over-density of stars is noticeable, which corresponds to the GC Ter 8 \citep[which has a spectroscopic $\feh = -2.3$,][]{2014MNRAS.443.1425S}. 
\subsection{Metallicity gradient}\label{metgrad}

\begin{figure}
\includegraphics[width=0.49\textwidth]{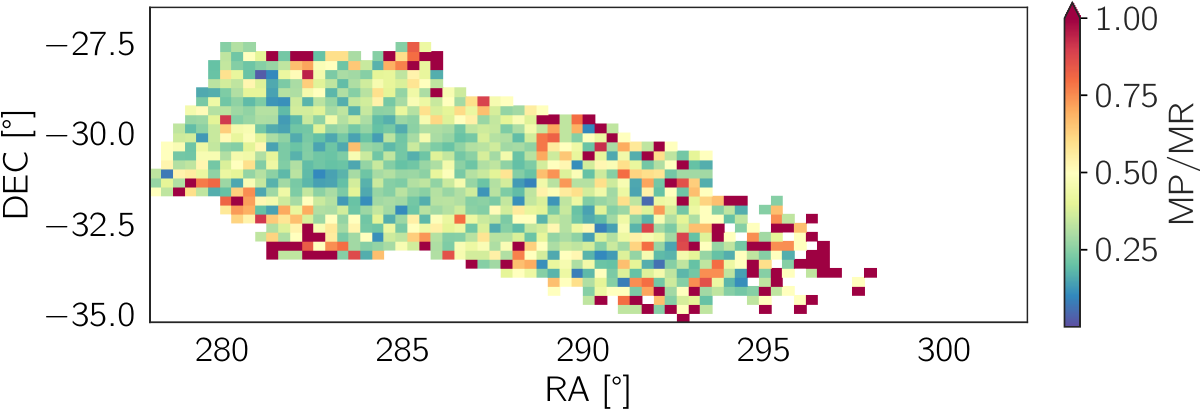}
\caption{2D histogram of the ratio between our final selected MP ($\feh < -1.3$) and MR ($\feh > -1.0$) populations. The binning and the colour-coding are the same as the ones of Figure \ref{MPMR}.}
\label{new_gradient} 
\end{figure}

Many studies \citep[e.g.][]{2010ApJ...720..940K, 2015ApJ...805..189H, 2017A&A...605A..46M, 2020ApJ...889...63H, 2021A&A...654A..23G} have shown that a metallicity gradient is present both in the streams and in a small central region of the Sgr remnant, which is connected to the intricate chemo-dynamical evolution of this dwarf galaxy. The PIGS data is an excellent dataset to study the metallicity gradient in the core of the galaxy. In Figure~\ref{new_gradient} we show the same 2D histogram as in Figure~\ref{MPMR}, this time with our MP ($\feh < -1.3$) and MR ($\feh > -1.0$) populations from the previous section, showing again the MP stars dominating at the edges and the MR in the central region.

\begin{figure}
     \begin{subfigure}[b]{0.44\textwidth}
         \includegraphics[width=\textwidth]{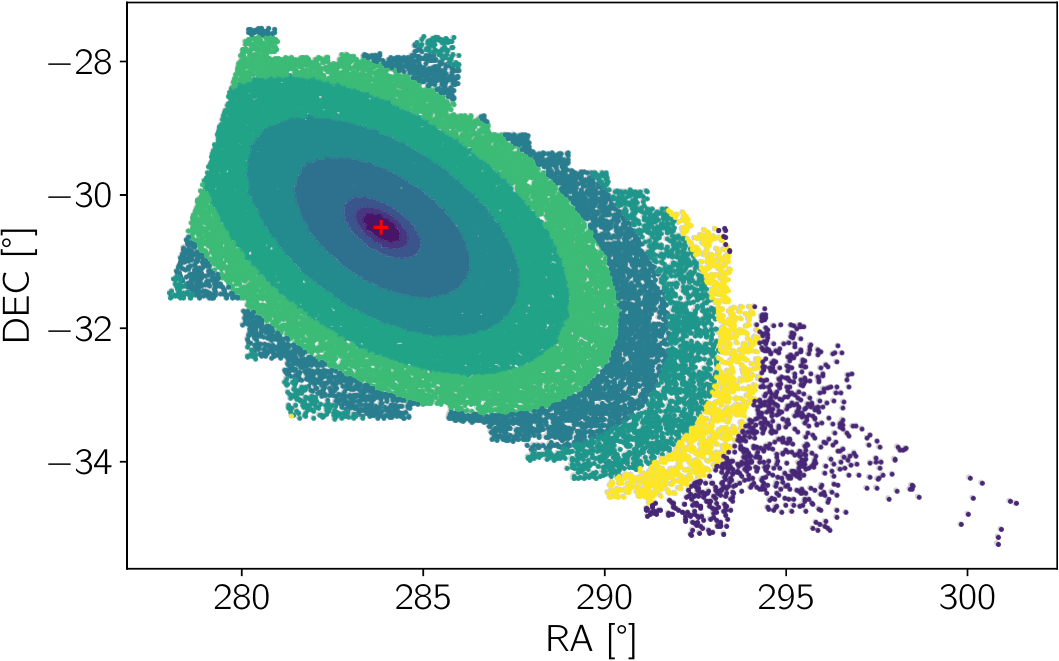}
     \end{subfigure}
     \hfill
     \begin{subfigure}[b]{0.44\textwidth}
       %  \centering
         \includegraphics[width=\textwidth]{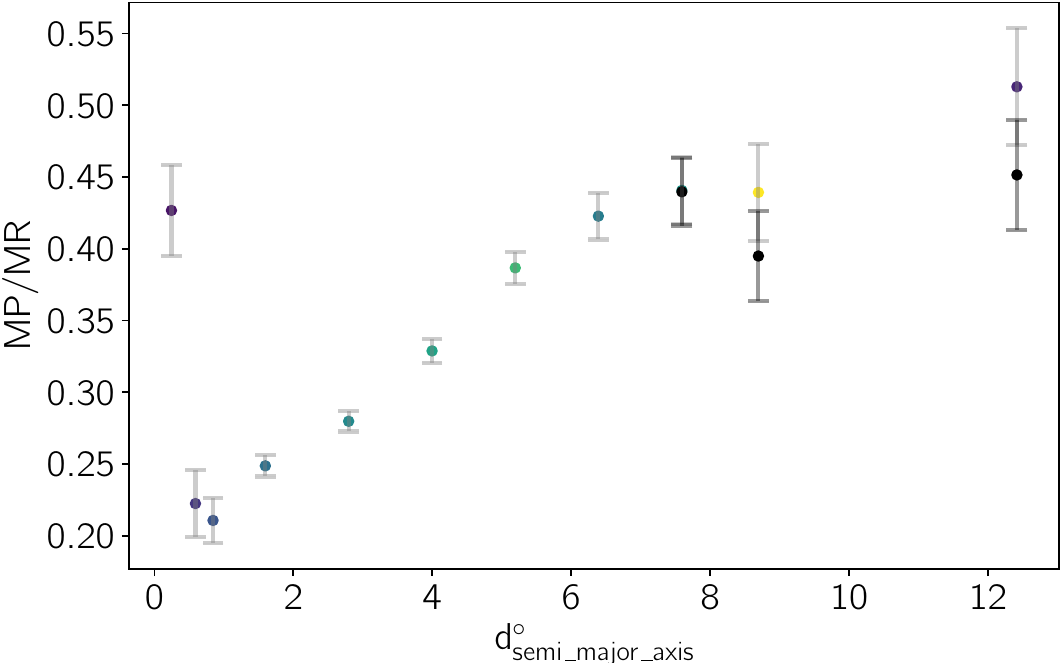}
    \end{subfigure}
     \caption{\textbf{Top:} Division of our Sgr sample in concentric ellipses. The parametrisation of each ellipse is defined using the fitted parameters from the MR population (see Table \ref{tab:table-name}). The centre of each ellipse is represented by the red cross ($283.830^{\circ}$ and $-30.493^{\circ}$). 
     \textbf{Bottom:} The ratios of MP/MR stars in each ellipse. The error bars are calculated assuming a Poissonian distribution and the colours of the points correspond to the colours in the top panel. The distances on the x-axis are calculated along the semi major-axis of each ellipse with respect to the centre, taking the middle distance between two consecutive ellipses. For the outer three ellipses, the black points represent the ratios after removing the globular clusters Arp2 and Ter8. }
\label{rings}
\end{figure}

Inspired by the idea of \citet{2017A&A...605A..46M} who mapped the change of metallicity as a function of the projected distance from the Sgr centre, we divide Sgr in concentric ellipses to examine the MP/MR ratio for each section, see Figure \ref{rings}. The ellipses are built using the parameters that we derived for the MR population in Section~\ref{model_sec}, which appear in Table \ref{tab:table-name}. The ellipses have a fixed width of 1.2 deg along the semi-major axis, except for the innermost region, where we chose smaller ellipses to probe the effect of M54 (with each bin containing at least 300 stars), and for the two outermost bins, where the density of Sgr stars drops rapidly. The area covered is significantly larger compared to the one considered by \citet{2017A&A...605A..46M}, as we computed the ratio out to $ \sim 12^{\circ}$ from the position of M54 along the Sgr semi-major axis, whereas the previous analysis only went out to $\sim 0.15^{\circ}$ (or 9'). Similarly, also the number of targets investigated is greatly increased with respect to the spectroscopic sample of \citet{2017A&A...605A..46M}, which accounted for 235 stars. 

By computing the ratio MP/MR for each division, as shown in Figure~\ref{rings}, it is possible to appreciate the change of the metallicity moving away from the centre. The distances are set along the semi-major axis of each ring with respect to the centre of the MR population (with $\feh > -1.0$, see Table \ref{tab:table-name}) using the projected coordinates from equation \ref{eq:xy}. The error bars are calculated assuming a Poissonian distribution. We find that Sagittarius presents a clear negative metallicity gradient -- the relative number of metal-poor stars is higher at larger radii from the centre of the galaxy. 

For the central ellipse, the stellar budget of M54 contributes by enhancing the relative number of metal-poor stars. The last rings, located furthest away from the centre of the galaxy, have higher uncertainties due to the lower stellar density in the outer Sgr region. The outer two ellipses also contain the two metal-poor globular clusters Ter8 and Arp2, with \feh $\sim -2.37$ and \feh $\sim -1.77$ \citep[][]{2008AJ....136..614M}, at $(\alpha, \delta) \approx (292^\circ,-34^\circ)$ and $(\alpha, \delta) \approx (292^\circ, -30^\circ)$ \citep{2010AJ....140.1830G}, respectively. For the outer three ellipses, the black points illustrate the MP/MR ratios after removing the stellar contribution from these two clusters. Excluding these two systems, we find that the trend seems to flatten starting at $d \sim 8^\circ$.

\section{Discussion} \label{discussion}
\subsection{Sources of uncertainties}
Our analysis is subject to uncertainties, which we discuss below. Overall, they do not significantly affect our main conclusions. 

\subsubsection{MW contamination}
\begin{figure}
     \centering\begin{subfigure}[b]{0.33\textwidth}
         \includegraphics[width=\textwidth]{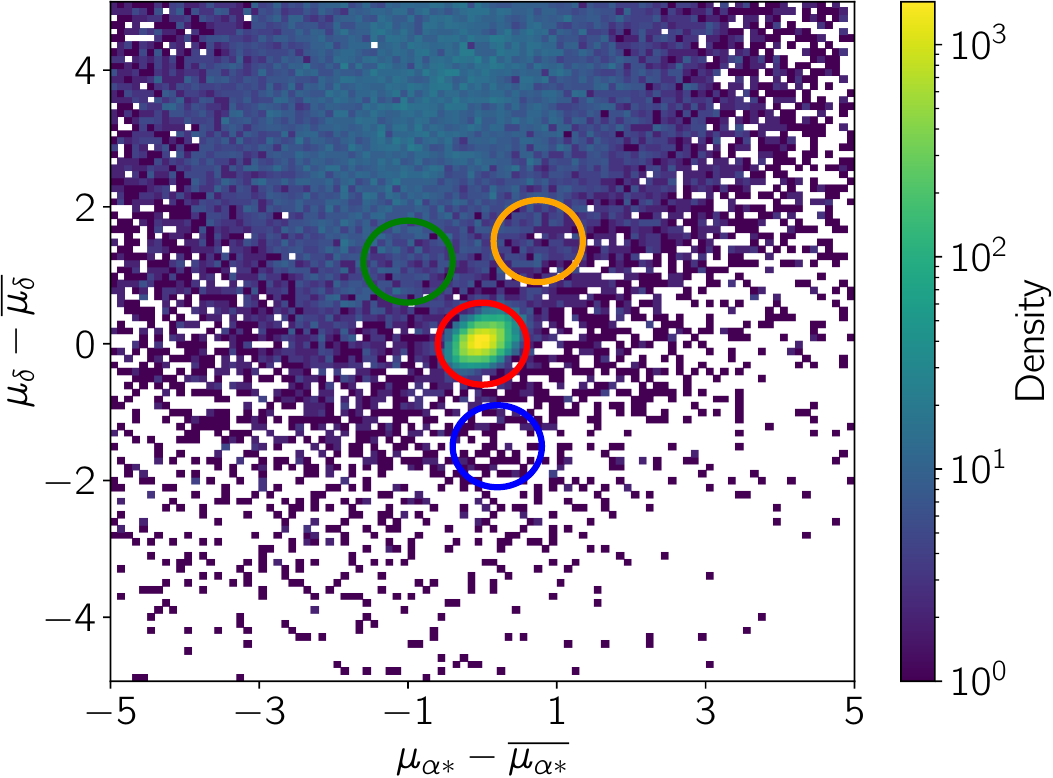}
     \end{subfigure}
     \vspace{0.12cm}
     
     \begin{subfigure}[b]{0.43\textwidth}
         \centering
         \includegraphics[width=\textwidth]{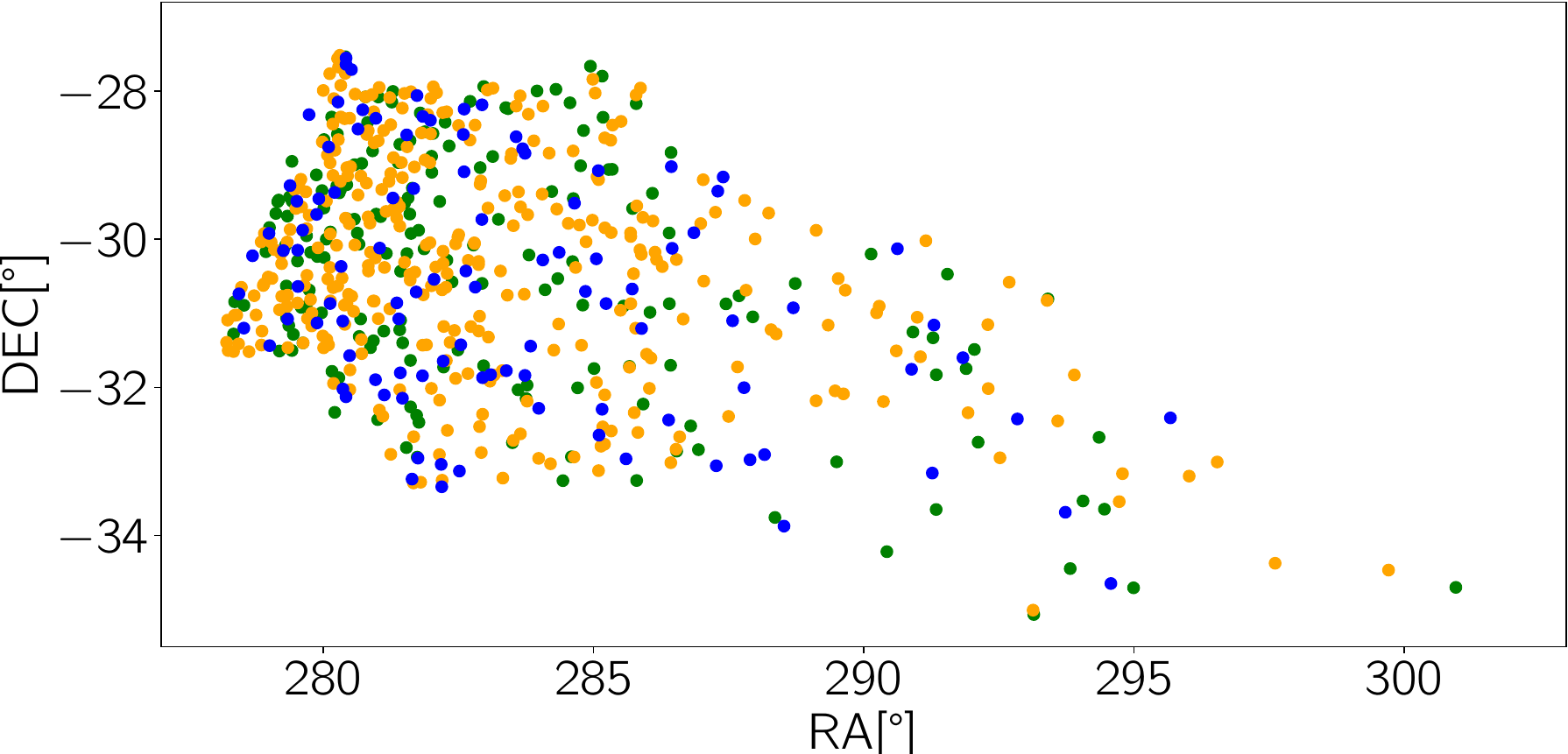}
     \end{subfigure}
     \vspace{0.12cm}
     
     \begin{subfigure}[b]{0.44\textwidth}
         \centering
         \includegraphics[width=\textwidth]{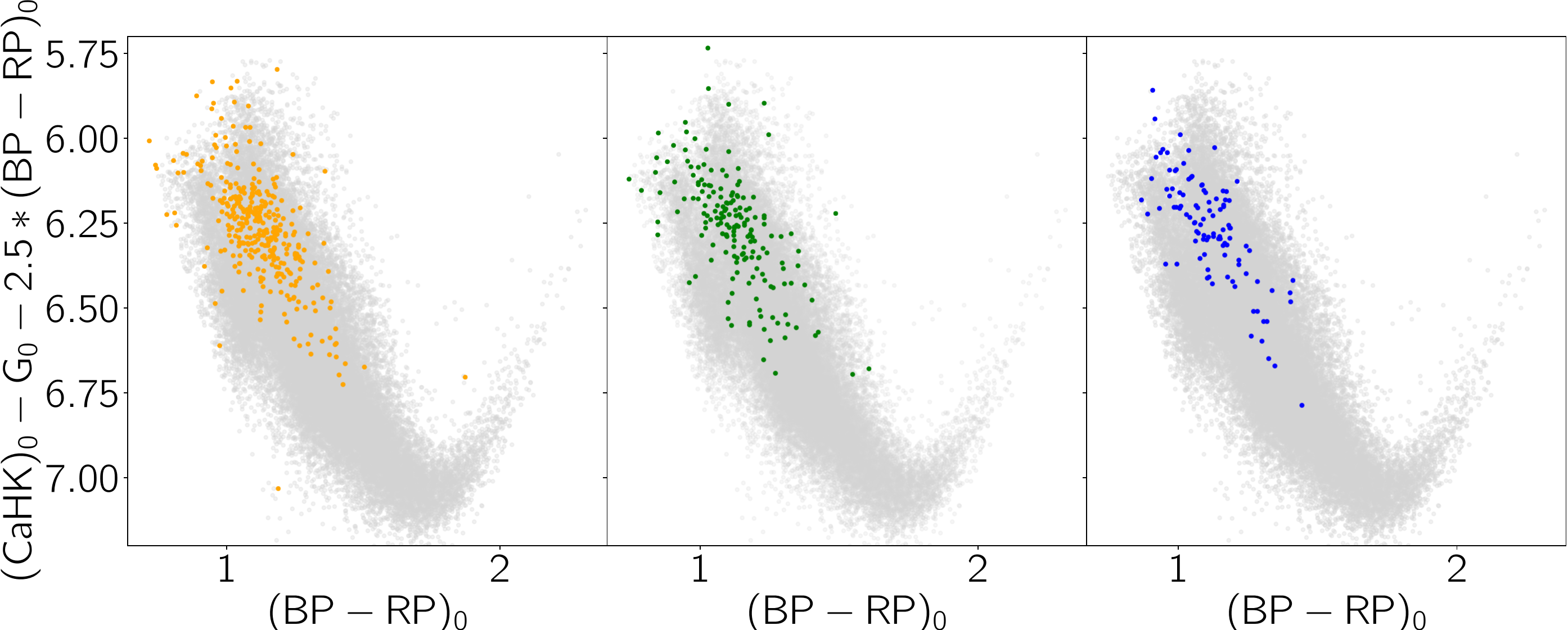}
     \end{subfigure}
     \caption{\textbf{Top:} PM distribution of stars in our Sgr sample before making a proper motion cut (the same as in Figure \ref{density}), with our selected Sgr sample in red, and three other indicated regions as MW control fields. \textbf{Middle:} RA/Dec distribution of the stars from the control regions, in corresponding colours to the circles above. \textbf{Bottom:} Colour-colour diagrams of our final Sgr selection (grey) and the stars in the control regions on top, again in the same colour coding.}
\label{cont}
\end{figure}

Despite the numerous cuts applied, some Milky Way foreground (or background) stars could still remain in the Sgr sample. 
We tested the level of contamination by selecting some Milky Way control regions in different parts of the proper motion space, to check their numbers and to see how they are distributed in the footprint and the colour-colour diagrams. We apply exactly the same cuts as to our Sgr sample, except for the proper motions. We selected three circular regions with the same PM radius as for our Sgr selection in roughly the same region of the PM space,
which are shown in orange, green and blue circle in Figure~\ref{cont}. Two of these fields, the green and orange circles, have higher densities than we expect in the Sgr region and show a pessimistic case of what the contamination level could be (175 and 332 stars). The control field depicted in blue contains even fewer stars (only 100). These numbers are to be compared with the 44\,785 stars in our Sgr selection. From the numbers of stars in each control region, it is possible to compute the percentage of the possible level of contamination left, which reaches at maximum $\approx 0.4\%, 0.7\%\, \mathrm{and}\,0.2 \%$ for each field, respectively.

We show their distribution in RA and Dec in the middle panel of Figure~\ref{cont}. The stars are mostly concentrated closer to the Galactic plane (on the left), and their density fades away towards higher RA (further away from the plane). The low ratio of MW stars left in the outer part of Sgr indicates that the contribution from the metal-poor MW halo is not dominant and it should not affect our metallicity analysis in the lower density regions. 

We inspected the location of the control fields in the colour-colour diagram (bottom panels) and found out that they are mostly located in the bluer part of the diagram, overlapping with the region that we identified as Sgr red clump stars. The contamination appears to be split between the MR and MP groups in roughly equal proportions compared to the Sgr stars, therefore they should not bias the results from our metallicity analysis. We tested whether it made a difference to our main results to exclude the RC region of the colour-colour diagram in our Sgr selection (using the blue line in Figure~\ref{panel_2}). We found no large differences, and therefore decided to keep the red clump region in our analysis.  

As described in Section \ref{model_sec}, we considered the MW contribution in the model fitting and found that it was unconstrained and did not impact the main results. This is consistent with our estimate of the MW contamination in this section, finding that it is very small. 

\subsubsection{Brightness cut}

The choice of the magnitude cut ($G_{0} = 17.3$), which noticeably reduced the Sgr sample (80\% of the original sample within the available magnitude range), has the advantage to give us a clean sample in the sense that the astrometry and the photometry are better constrained enabling a more effective cleaning from the MW foreground stars. Whilst fainter targets might be interesting to have a more stars of this galaxy, we decided to reject the horizontal branch and the red clump region because it would make the selection of MR and MP populations more complicated and less complete. Furthermore, discarding the horizontal branch means to remove variable stars which start appearing in this region of the diagram.

\subsubsection{Iso-metallicity lines} 

For the derivation of the iso-metallicity lines, we considered using the \CaHK and \Gaia G, BP and RP uncertainties. The uncertainties in \CaHK are all $<0.08$, and when included in the fitting operations our results did not change significantly. The errors on the \Gaia photometry are much smaller than the \CaHK, hence we ignored them as well. As mentioned earlier in the text (Section~\ref{isolines}), the iso-metallicity lines stop to be trustworthy for the coolest stars, namely $\bprp > 1.6$, because in this region the  metal-poor and metal-rich sequences start crossing each other. For this reason we relied on the APOGEE \feh values to classify stars with $\bprp > 1.6$ as all being metal-rich. We also found that in the spectroscopic training sample, the location of stars in the \Pristine colour-colour diagram depends on the alpha abundances for the coolest/reddest stars (with $\bprp > 1.5$), resulting in a degeneracy between metallicities and alpha abundances. The alpha abundances in Sgr are different compared to those in the Milky Way, therefore the iso-metallicity lines derived from the training sample could be partly inappropriate for Sgr. This may be connected to our finding that the \CaHK shift between the spectroscopic training sample and the Sgr spectroscopy sample appears to depend on the metallicity of the Sgr stars used.

It is worth mentioning that the iso-metallicity lines were derived from a sample with low extinction towards the Galactic halo, whereas the Sgr region presents higher extinction since it is relatively close to the Galactic disc. Uncertainties in the extinction correction mean that this difference is expected to slightly increase the uncertainties in the metallicities for the Sgr sample, but it is not expected not create a systematic offset in the metallicity calibration.
\begin{figure}
     \begin{subfigure}[b]{0.44\textwidth}
         \includegraphics[width=\textwidth]{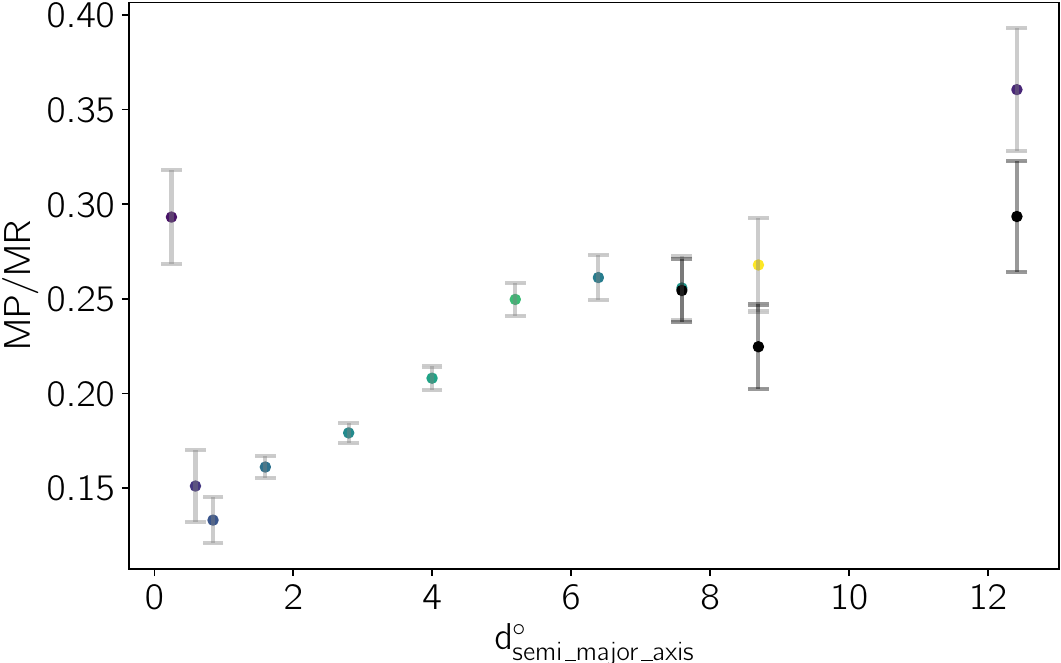}
     \end{subfigure}
     \vspace{0.12cm}
     
     \begin{subfigure}[b]{0.44\textwidth}
       %  \centering
         \includegraphics[width=\textwidth]{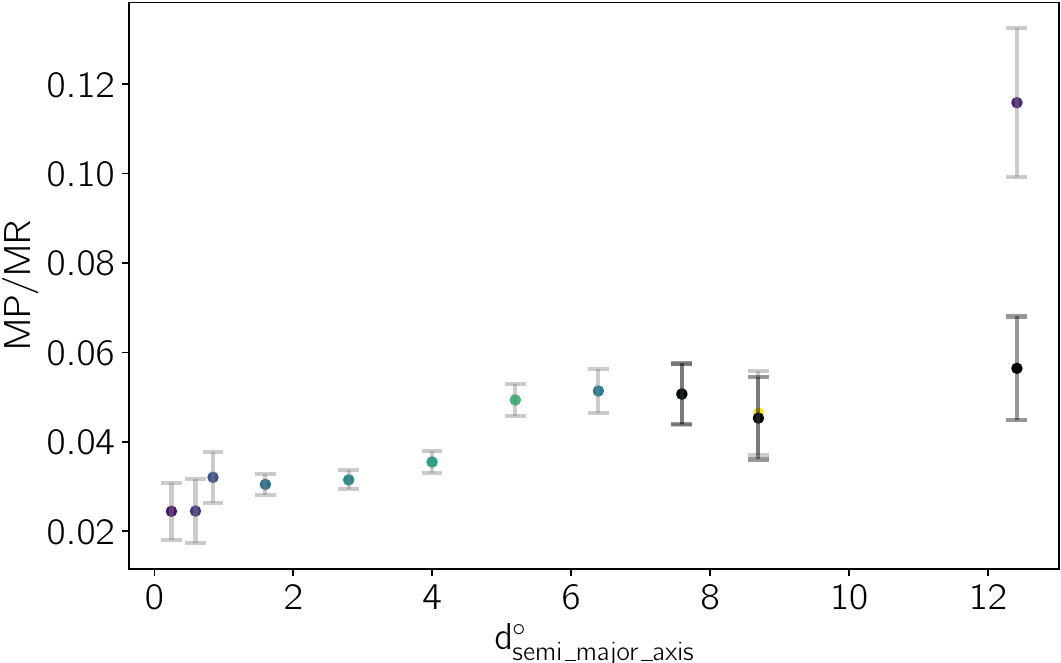}
    \end{subfigure}
     \caption{Same MP/MR ratio as in the bottom panel of Figure~\ref{rings}, but presenting populations selected with different photometric metallicity thresholds for the MP population (top: $\feh <-1.5$, bottom: $<-2.0$). The metal-rich (MR) group is always defined as $\feh > -1.0$.}
\label{boundaries}
\end{figure}
\subsubsection{Definitions of MP and MR populations} \label{unc}

Figures~\ref{MPMR} and~\ref{new_gradient} showed that, even when shifting the metallicity boundary for the MP and MR regions between $-1.0, -1.3, -1.5$ or $-2.0$, the fraction of MP stars relative to the MR ones always increases with distance from the centre of Sgr, therefore the metallicity gradient remains and does not depend on our exact definition of the MP population. This effect was quantified in the bottom panel of Figure \ref{rings} for our main MP ($\feh < -1.3$) and MR ($\feh > -1.0$) populations, presenting a clear negative metallicity gradient. Choosing the two more metal-poor boundaries introduced in section \ref{division} ($\feh < -1.5$ and $< -2.0$) for the MP population instead, we reproduce a similar trend of the $\mathrm{MP/MR}$ ratio rising away from the centre for these cases, presented in Figure~\ref{boundaries}. In both panels, it is possible to notice a flattening in the trend starting at around $d \sim 6^\circ$. The fraction of stars with \feh $< -1.5$ appears to be relatively constant in the outermost rings, except for the last annulus. There the MP/MR ratios do experience an important rise due to the presence of Ter8 and Arp2. From the same figures it can be observed that the flattening is stronger when the contribution from the two metal-poor clusters Arp2 and Ter8 is removed. For the $\feh < -2.0$ stars, there is no sign of M54 in the innermost rings anymore, and the rest of the trend shows a similar flattening as the previous case.  

It is beyond the scope of this paper to further quantify the gradient (e.g. in terms of dex/kpc), which requires individual metallicity estimates for all our stars. For that, a larger spectroscopic sample of Sgr stars is needed, either to be used by itself or to better constrain the photometric metallicities.

\subsubsection{Model fitting}
The exponential elliptical profile adopted in our model fitting is a relatively simple approximation. Although it roughly corresponds to the projected shape of Sgr, the real distribution of Sgr stars is more complex. The model fits may also be influenced by inhomogeneities throughout the footprint. Both could result in unrealistic uncertainties from the MCMC. We checked whether the results from the model fits depend on the binning of the data, and find no significant changes.

To investigate whether differences for the model parameters between the MP and MR populations are likely to be real, we performed model fits on Sgr stellar populations in smaller metallicity bins (0.2~dex wide). We found a trend with metallicity for the scale radius, but found no clear evidence of trends with metallicity for the eccentricity, position angle and centre of the models. We also concluded from this exercise that the uncertainties from the MCMC appear to be underestimated. In view of these considerations, the changes between the different model parameters (see Table~\ref{tab:table-name}), except for the scale radius, should be treated with caution. 

\subsection{Sgr metallicity gradient compared to the literature} \label{sect_gradient}

Evidence of a metallicity gradient in Sgr was already reported early on \cite{1999MNRAS.304..633B}. Those authors connected the presence of a metallicity gradient to a significant age variation in the dwarf system, caused by a protracted star formation history. \citet{2001A&A...377..389A} studied two Sgr fields of 2x2$^\circ$, discovering a variation of $-0.2$~dex along the Sgr major axis that was linked to an age variation inside the core. 

Thanks to near-infrared photometry, \citet{2013MNRAS.436..413M} studied the variation of the fraction of metal-poor ($-1.6 < \feh < -0.9$) and intermediate, metal-rich ($\feh > -0.7$) stars in an area of eleven square degrees and revealed that the metal-poor population is more spread throughout the dSph, while the metal-rich stars ($\feh \sim 0.0$) are grouped in ellipsoidal distribution around the centre of the galaxy. \citet{2013MNRAS.436..413M} identified clear traces of a metallicity gradient away from the region dominated by the bulge population (from RA $ \sim287^\circ$). 
\citet{2017A&A...605A..46M}, through their spectroscopic study of 235 stars in the Sgr core and in M54 (all within 9' of the centre), uncovered a metallicity gradient as well, finding a higher fraction of metal-rich stars in the centre. They speculated that the metallicity gradient in that region can be linked to an extended star formation history, in which recent metal-rich bursts took place in the central area of Sgr which might have caused later stellar generations to be more centrally concentrated. 

Our results are consistent with previous works, namely our metal-rich population (MR, $\feh > -1.0$) dominates the innermost region while our metal-poor population (MP, $\feh < -1.3$) becomes more important at larger radii, i.e., there is a metallicity gradient in our data. In line with these results, the greater value of the scale radius for the more metal-poor stars could be an indication that, generally, the stars belonging to this category are more smoothly distributed on larger distances compared to the more centrally concentrated metal-rich stars. By comparing the $r_{e}$ values derived for a number populations in narrower metallicity bins, we found a clear increase of this parameter moving to lower metallicities.

We note that our work covers an area of $\sim 100$ square degrees, extending to $\sim 12$ degrees from the very centre of the galaxy, which is considerably larger than the region covered in previous studies, which typically focused on the very inner part of Sgr. Here we find that the metallicity gradient of Sgr extends beyond the central part of the galaxy and it manifests all the way outwards to the stream. The Sgr streams have metallicity distributions peaking at lower metallicity values than its core, \feh $\approx -0.8/-1.1$ and \feh $\approx -0.5$ respectively \citep{2020ApJ...889...63H}. Our analysis revealing a higher fraction of stars with $\feh < -1.3$ in the outskirts (RA $> 290^{\circ}$) is in congruence with these findings, supporting the scenario in which the most metal-poor stars were the least bound to Sgr, and have been tidally stripped away from the core first.

This result highlights the power of the \Pristine data, which enables us to characterise not only the dense central regions of the dwarf galaxy but also the outskirts. Combined with \textit{Gaia} astrometric information, the photometric metallicity information is ideal to study the structural properties of a dwarf system along its entire extension.

\subsection{The formation and evolution of Sgr}

At this stage of the work, it is tantalising to tie the results from our metallicity analysis to the history and evolution of Sgr. In general, the morphology and the star formation history of a system can be heavily influenced by various physical processes that can be both internal, such as feedback from SNe events and gas pressure support, or triggered by external factors, e.g. ram pressure stripping and tidal disturbances caused by Galactic tides (tidal stripping and tidal stirring) \citep[][]{ 2001ApJ...559..754M,2006MNRAS.369.1021M, 2010ApJ...725.1516L} or mergers \citep{2015A&A...575A..59S,2016MNRAS.456.1185B}. 

Recently, \cite{2018MNRAS.478.5263T} and \cite{2020MNRAS.497.4162V} in their simulations assumed Sgr to have been a gas-bearing dwarf spheroidal galaxy before it fell into the Milky Way. According to this theory, Sgr has conserved its origin of dSph despite the tidal interaction with the Milky Way, and is predicted to dissolve over the next Gyr \citep{2020MNRAS.497.4162V}. Others suggest that the Sgr progenitor was a gas-rich, flattened rotating system that transformed into a dSph due to tidal stirring in the interaction with the Milky Way, and whose inner core might survive the next pericenter passage \citep{2010MNRAS.408L..26P, 2010ApJ...725.1516L, 2021ApJ...908..244D}. \cite{2010ApJ...725.1516L} suggested that the Sgr progenitor resembled the Large Magellanic Cloud and they described it as a disky galaxy whose stellar populations formed a bar-like structure that survived until the second pericenter passage. The simulations of \cite{2022ApJ...932L..14O} suggest that Sgr hosted a rotating component that has been perturbed during the interaction with the MW, but that this component was not the dominant fraction of the stellar mass. There is no consensus yet on the progenitor of Sgr, but more data might help to reach a conclusion. The arguments about the nature of the Sgr progenitor and its subsequent evolution are mainly based on the kinematical properties of Sgr and its stream, but there is another dimension as well: the chemistry.

\subsubsection{Processes shaping the metallicity gradient}

One process which is known to shape the age/metallicity gradients in satellite galaxies is ram-pressure stripping, which is responsible for removing the gas that was originally in the dwarf galaxy \citep{2001ApJ...559..754M,2006MNRAS.369.1021M}. Ram-pressure stripping first removes the gas in the less dense outer regions of a dwarf galaxy, and removes the central gas reservoir at the very end. This means that new stars could be forming for longer periods of time in the centre compared to the outer regions. The gas in the inner regions is chemically enriched due to the prolonged star formation, hence this process can lead to age and metallicity gradients. We know that the Sgr core contains a population of relatively young stars ($<2$~Gyr, e.g. \citealt{2007ApJ...667L..57S}), so it must still have had gas up to those times. However, no gas has currently been detected in Sgr \citep{1999A&A...349....7B, 1994MNRAS.270L..43K}. According to the modelling of \cite{2018MNRAS.478.5263T}, $30-50$ per cent of the gas in Sgr was stripped $\sim 2.7$~Gyr ago (the first time it crossed the Galactic disc), while the complete loss of the remaining gas took place $\sim 1$~Gyr ago, when it crossed the disc the last time. Another process related to metallicity gradients in satellite galaxies is tidal stripping due to interaction with the host galaxy. Tidal stripping affects the more diffuse component of a dwarf more strongly, because it is less tightly bound to the galaxy. In Sgr, tidal stripping has led to the removal of much of its stellar content, which is now forming the large Sgr stream. The Sgr streams are more metal-poor than the progenitor. The metallicities of their stellar populations span between \feh $\sim -2.5$ and $\sim -0.5$ \citep{2015MNRAS.451.3489D}. This indicates that the metal-poor stars were the ones that were less bound to the core, suggesting a radial metallicity gradient in Sgr. We also found in this work, using the radial-velocities available from the spectroscopic catalogues, that the velocity dispersion in the core of Sgr is higher for metal-poor stars than for metal-rich stars, supporting this scenario.

It has been shown that there is a difference between the metallicity gradient detected in dSph and dwarf irregular galaxies. Dwarf irregular galaxies (dIrrs) are rotationally supported and have a disky dwarf progenitor, while dwarf spheroidals are recognised to be pressure supported systems. Generally, the first category shows a steeper decreasing gradient profile with respect to the flatter trend present in dSph systems \citep{2001ApJ...559..754M,2022arXiv220608988T}. Since the progenitor of Sgr is still under debate, we can expect that depending on the assumed scenario for the progenitor - a dSph or disky rotating system -  the metallicity gradient would be less or more pronounced. It is also necessary to bear in mind that the transition from a disk galaxy to a dSph can be caused by tidal stirring process, which have been invoked in the case of Sgr for being responsible for its observed elliptical shape \citep{2010ApJ...725.1516L}.

\subsubsection{Connecting ages and metallicities}

The relation between metallicity and age gradients is important in constraining the processes behind the metallicity gradient. Many works showed that age-metallicity relations can be derived for red giant branch (RGB) stars in dwarf galaxies using information from their SFHs and CMDs \citep{2011AJ....142...61C,2015MNRAS.454.3996D,2017MNRAS.469.4999D}. However, these strategies are effective when the SFH concerns the same region of interest of the stars. We did not directly associate the metallicity gradient in our work with a possible age gradient as it would have been beyond the scope of this paper. We could assume age-metallicity relations for Sgr from other works \citep[see for instance][]{2000AJ....119.1760L,2006A&A...446L...1B, 2007ApJ...667L..57S}, but this is not trivial as most of them analyse fields of only a few degrees in the very central part around M54. Keeping that caveat in mind, if we adopt the age-metallicity relation presented by \cite{2007ApJ...667L..57S}, we can speculate that our identified MR and MP populations have ages of $\sim 4-8$ and $\gtrsim 10$ Gyr, respectively. 

Another way to connect ages and metallicities is by using their alpha abundances. The knee in the $\feh - \alphafe$ diagram indicates the switch from star formation on shorter time-scales, where the SN type II were the main contributors to the gas enrichment, to star formation on longer time scales, during which the metal enrichment from SN type Ia started to contribute significantly. The knee in the $\feh - \alphafe$ diagram in M54 and the Sgr stream has been derived to be located at $\feh \sim -1.3$ \citep{2010A&A...520A..95C,2014MNRAS.443..658D}, and has been connected to an age of $\approx 11 \mathrm{Gyr}$. The knee is still somewhat unconstrained in the core and could be different, since the data suggests it experienced a prolonged and complex star formation history. However, assuming it is around the same metallicity, this is another indication that our MP population is significantly older than the MR population.

Sgr has interacted with the Milky Way for $\gtrsim 8$~Gyr, and the first close pericenter passage of Sgr is predicted to have happened $\sim5-6.5$~Gyr ago \citep{2010ApJ...714..229L,2020NatAs...4..965R}. Given an age of $\sim 4-8$~Gyr for the MR stars, this means that the younger metal-rich stars might have been born during the first encounter as a consequence of the infalling triggered by the tidal interaction. Also the older metal-rich stars and the metal-poor stars would already be present at the time of the first interaction. However, we can not exclude that these populations will have been affected differently by the tidal interaction with the Milky Way, depending on their internal properties at the moment of infall -- i.e. how tightly they were bound to the remnant, and whether they had any rotation or not.

\subsubsection{Metallicity gradient and its interpretation in Sgr}

The processes playing a role in the formation of metallicity gradients in low-mass dwarf systems ($\mathrm{M_{*}}\lesssim 10^{8}-10^{10}\,\mathrm{M_{\odot}}$) remain not fully understood. There is not a clear consensus whether metallicity gradients are formed by protracted central star formation episodes with respect the outer regions, or if they are driven by processes acting more on the older, more metal-poor stars by moving them outwards \citep{2018A&A...616A..96R, 2022arXiv220608988T}, and/or a combination of these.

The complex combination of internal (such as rotation, orbits, angular momentum content) and environmental factors (tidal and ram-pressure stripping) plays a role in shaping and weakening metallicity gradients, and it is difficult to disentangle these additional factors from the role of an extended SFH \citep{2006MNRAS.369.1021M,2010AdAst2010E..18S,2022arXiv220608988T} and consequently link the gradient to a progenitor. 

By looking at our figures illustrating the various [MP/MR] ratios, the fact that it is still possible to detect a gradient might be interpreted as a hint of disky progenitor in which the gradient should have been strong enough to be partially preserve until today. The change of steepness observable at RA $\sim 290^\circ$ might be related to the transition from the outer core to the stellar stream, which might mitigate the profile observed for the very inner part where the various star formation bursts took place. 

The work of \cite{2021MNRAS.501.5121M} reported that a late gas accretion is a further event that can weaken or flatten an existing metallicity gradient in a dwarf galaxy. If we consider that Sgr might have experienced a first encounter with the MW around 5-6 Gyr ago, this factor should be added to the secular processes which act in weakening the trace of the original gradient. If we add to this picture the protracted SFH in Sgr, according to \cite{2016MNRAS.456.1185B}, a steep gradient can only be present in the case of a past merger event, responsible of scattering the old metal-poor component, followed by an \textit{in situ} metal-rich star formation from the infalling central gas.

It is however ambitious to derive robust conclusions about the progenitor of Sgr before it fell into the MW -- whether it was a rotating disky galaxy or a pressure supported spheroidal galaxy -- without ages or individual metallicities. We previously discussed how ages could help in getting a more complete picture of the evolution of Sgr. On the other hand, individual metallicities would give the possibility of quantifying the slope of the gradient and spotting possible brusque changes in its trend that we are not able to detect with our metallicity division. If we were able to quantify the radial metallicity gradient, we could compare it with similar pressure-supported dSphs but located in isolated environments. This comparison would enable us to evaluate the impact of both internal feedback and Sgr properties (such as mass and angular momentum) and exterior mechanisms on the Sgr metallicity gradient.

In our fits of the the spatial distributions of the MP and MR populations, the only significant difference we found was that of the scale radius, with the MR population being more centrally concentrated than the MP population. 
This reflects their spatial distribution and the detected trend in the [MP/MR] ratio: a younger, more centrally concentrated metal-rich component is surrounded by an older, more disperse metal-poor population present at increasing radii. To further disentangle the different processes which could have shaped the metallicity gradient in Sgr, the spatial properties should be accompanied by age information. 

Besides the change in the scale radius between the MP and MR populations, if the other small differences we found in the structural parameters from the model fitting were real, what could they tell us? We found a small shift between the centres of the MR and MP populations in the RA direction. This could hint that the tidal interaction might have also played a more severe role in shaping and shifting the extended MP population compared to the MR population, which formed on a longer time-scale from the central gas reservoir. The change of the position angle for different metallicity populations could be related to the fact that they have interacted with the Milky Way tidal field over different periods of time and thus their orbits and positions have evolved differently. It could also be due to rotation within the dwarf galaxy, which may be different for the young (metal-rich) and the old (metal-poor) populations, as suggested in for example \citet{1997AJ....113..634I}. The values of the ellipticity for both populations (0.566 and 0.592 for MP and MR respectively) are lower than the ellipticity of $\sim 0.65$ presented by \citet{2003ApJ...599.1082M}. But it is still high, which is a sign of the process of tidal elongation of Sgr induced by interactions with the Milky Way. The higher \textit{e} value for the MR population might also be related to its lower velocity dispersion.

\subsubsection{Comparison with other dSph}
The Fornax dSph is an interesting example of a dwarf galaxy that shows some similarities with the Sgr dSph galaxy. The work of \citet{2012A&A...544A..73D} reports the existence of a radial gradient of age and metallicity, with more metal-rich and younger star forming episodes condensed in the central region of the system, and oldest and more metal-poor stars (with an age of $\geq 11 \mathrm{Gyr}$) appearing at all radii.
The dominance of the intermediate-age stellar populations and the protracted SFH for both systems indicate that their dynamical masses were sufficient to retain enough gas (before the complete gas loss) to keep forming stars in their inner regions, after the gas fell back into the central potential well. It is also suggested that the cause of repeated peak in the star formation might be a merger with a gas-rich companion. In the case of Fornax, a precise chemical estimation predicts the alpha-knee to occur at \feh $\sim -1.5$ (corresponding to an age of 7-10 Gyr). \cite{2006A&A...459..423B} found the older, more metal-poor population ($\mathrm{[Fe/H]} < -1.3$, age $> 10$ Gyr) to be more spatially extended than the more metal-rich and younger population (with $\mathrm{[Fe/H]} > -1.3$ and ages between 2-8 Gyr), which was more centrally concentrated. Comparable results are presented in other works, such as \cite{1998PASP..110..533S} and \cite{2019ApJ...881..118W}. Also \cite{2008ApJ...685..933C} and \cite{2013MNRAS.433.1505D} reported a protracted SFH and detected a gradient in the stellar populations of which the youngest and most metal-rich formed more segregated in the centre of the galaxy.

Another compelling example is the Sculptor dSph galaxy, in which the metal-poor ($\mathrm{[Fe/H]} < -1.7$) and metal-rich ($\mathrm{[Fe/H]} > -1.7$) stellar components posses different kinematics and spatial distributions. I.e. the more extended metal deficient population has higher velocity dispersion than the metal-rich population \citep{2004ApJ...617L.119T}, which was likely created after the enriched original gas sank back to the centre.

\subsection{New sample of very metal-poor stars in Sgr}

Very and extremely metal-poor stars in dwarf galaxies strongly reflect the early star formation history of their host systems and, being tracers for the first nucleosynthesis events, can help to constrain the properties of the first stars, such as their initial mass function. The high-resolution study targeting the metal-poor tail of Sgr performed by \citet{2018ApJ...855...83H} discovered a similarity in the chemical composition between Sgr and the MW halo for these iron-depleted candidates, hinting that galaxies like Sgr contributed to the MW stellar halo. \citet{2017ApJ...845..162H} reached a similar conclusion. By studying this type of objects, focusing on their chemical composition, it is possible to study the past and/or ongoing accretion events.

Despite the lack of abundances measurement, the distribution of our unprecedented selection of VMP stars opens a window into the star formation processes behind this ancient population. It suggests that no such stars were formed recently in the inner area, as we found them to be quite diffuse. Yet they do also show a higher density (similarly to the other populations) in the centre, as seen in Figure \ref{VMP}. The distribution of this population can be linked to the gradual disruption of the progenitor, now leaving a core and the wide stellar streams. Indeed, it is also likely that tidal impulses, which Sgr experienced from the MW during passages at its pericenter, provoked a violent mixing of stars of different populations. This could have erased a possibly more pronounced radial metallicity profile in the Sgr progenitor, which was suspected to show an even greater fraction of more metal-rich stars tightly bound in the interior regions \citep{2007ApJ...670..346C}. These stars would have been removed at later times compared to the older and more metal-poor population, creating the known metallicity variations along the Sgr stellar streams \citep{2007ApJ...670..346C,2020ApJ...889...63H}. Within this perspective the remaining VMP objects in the core can be seen as left-overs of the more ancient population ($\gtrsim 10 $ Gyr), once hosted in the galaxy progenitor, which has been gradually stripped away and deposited in the streams, known to be on average 1 dex more metal-poor than the core \citep{2015MNRAS.451.3489D}. 

There is no overlap between our VMP selection and the APOGEE spectroscopic data (which contains mostly metal-rich stars), while the cross-match with the PIGS spectroscopic catalogue reveals 115 stars in common.  Additional spectroscopic follow-up spectroscopy of the VMP candidates in our sample is required to further study the nature of these stars and the early chemical evolution of Sgr, and this effort is on-going. 

\section{Conclusions and Future work}\label{conclusions}

In this work we presented the largest photometric metallicity study of the core of the Sagittarius galaxy to date, using metallicity-sensitive narrow-band photometry from the Pristine Inner Galaxy Survey (PIGS). To summarise the results:

\begin{itemize}
    \item By combining the PIGS photometry with the precise astrometry and broad-band photometry from \textit{Gaia} EDR3, we were able to isolate bright giant Sgr stars ($G_{0}\leq 17.3$) to build an unprecedented sample of 44\,785 reliable Sgr members with metallicity information. 
    
    \item Using photometric instead of spectroscopic metallicities allows a much more homogeneous analysis of Sgr populations of varying metallicity. The PIGS data cover $\sim 100$ square degrees of Sgr out to ~12 degrees along the semi-major axis from the centre (corresponding to $\sim 5.5$ kpc at the distance of Sgr), covering most of remnant of the dwarf galaxy core. We divided the Sgr stars into different metallicity populations, with our two main samples being the metal-poor (MP) having \feh$<-1.3$ and the metal-rich (MR) having \feh$>-1.0$. 
    
    \item Our data reveal a metallicity gradient, with the metal-rich stars dominating in the inner regions and the metal-poor stars towards the outer regions. This is consistent with previous evidence of a metallicity gradient in Sgr, but we extend it to much larger radii than previously observed. 
    
    \item We fitted models of the stellar density distributions for populations with various metallicities, separating metal-poor and metal-rich stars at $\feh = -2.0, -1.5, -1.3$ and $-1.0$. The most striking difference we find is a change in the scale radius as function of metallicity, where the metal-rich stars are more centrally concentrated, while the more metal-poor component is more diffuse and distributed as a spheroid with a larger effective radius.
    
    \item The PIGS photometry is still sensitive to metallicity for very metal-poor (VMP, $\feh < -2.0$) stars. We previously used it to select stars for low-/medium-resolution spectroscopic follow-up, resulting in the largest sample of 100 spectroscopically confirmed Sgr VMP stars. In this work, we further used the PIGS photometry to build an unprecedented sample of 1150 VMP candidates in Sgr with $G_0 < 17.3$. This remarkable sample of iron-depleted stars is left over from an ancient population that was once hosted in the Sgr progenitor, which has likely partially been removed and distributed to the Sgr streams and/or the Galactic halo.
    
    \item We discussed how the history and evolution of Sgr could have impacted the various Sgr stellar populations. Our results are consistent with an outside-in quenching process with an older, diffuse metal-poor stellar population and a younger, more centrally concentrated metal-rich counterpart forming at later times. Sgr had an extended and rich star formation history, forming different stellar populations with a different spatial and chemical evolution. To further connect our detected metallicity gradient with the properties of the underlying stellar populations, we need a better precision in metallicity and, currently missing, information about the ages of different metal-poor populations. 
 
\end{itemize}

Spectroscopic studies on elemental abundances of dwarf systems pave the way in revealing their assembly and evolution histories in more depth, and increase the knowledge about metallicity distributions and age gradients in dwarf galaxies. For example, a strong connection exists between the initial mass function and the early chemical evolution. Chemical information allows to shed light on important aspects in the life of a dwarf galaxy, such as the frequency of star formation episodes, the stellar yields, and the mixing processes that took place in the interstellar medium. Another factor in the evolution of a dwarf galaxy is its interactions with other galaxies, and Sgr is a unique example of a complex disrupting dwarf system (core and streams). 

By exploring with high-resolution spectroscopy the chemical composition of the Sgr stellar populations, especially focusing on its metal poor tail, it will be possible to constrain its star formation history and characterise the early evolution of this dSph galaxy. We are planning spectroscopic follow-up of our unprecedented Sgr sample with metallicity information, the result of the powerful combination of the metallicity-sensitive photometry from the \Pristine survey \citep{2017MNRAS.471.2587S} and the excellent \textit{Gaia} EDR3 data.

\section*{Acknowledgements}

   We would like to thank the reviewer for carefully reading the manuscript and for giving many valuable suggestions.

    We thank the PIGS AAT observers Sven Buder, Geraint Lewis, Sarah Martell, Jeffrey Simpson, Zhen Wan and Daniel Zucker. 

	We thank the Australian Astronomical Observatory, which have made these observations possible. We acknowledge the traditional owners of the land on which the AAT stands, the Gamilaraay people, and pay our respects to elders past and present. Based on data obtained at Siding Spring Observatory (via programs S/2017B/01, A/2018A/01, OPTICON 2018B/029 and OPTICON 2019A/045, PI: A. Arentsen and A/2020A/11, PI: D. B. Zucker). 
	
	Based on observations obtained with MegaPrime/MegaCam, a joint project of CFHT and CEA/DAPNIA, at the Canada-France-Hawaii Telescope (CFHT) which is operated by the National Research Council (NRC) of Canada, the Institut National des Science de l'Univers of the Centre National de la Recherche Scientifique (CNRS) of France, and the University of Hawaii.
	
	AA, NFM and ZY gratefully acknowledge funding from the European Research Council (ERC) under the European Unions Horizon 2020 research and innovation programme (grant agreement No. 834148).
	
	ES acknowledges funding through VIDI grant "Pushing Galactic Archaeology to its limits" (with project number VI.Vidi.193.093) which is funded by the Dutch Research Council (NWO)
	
	PJ acknowledges financial support of FONDECYT Regular Grant Number 1200703. 
	
	GK, NFM, RI, VH and ZY gratefully acknowledge support from the French National Research Agency (ANR) funded project ``Pristine'' (ANR-18-CE31-0017).
 
  	JIGH acknowledges financial support from the Spanish Ministry of Science and Innovation (MICINN) project PID2020-117493GB-I00.
  	
  	D.A. acknowledges support from ERC Starting Grant NEFERTITI H2020/808240.

	This project has received funding from the European Union's Horizon 2020 research and innovation programme under grant agreement No 730890. This material reflects only the authors views and the Commission is not liable for any use that may be made of the information contained therein.

	The authors thank the International Space Science Institute, Bern, Switzerland for providing financial support and meeting facilities to the international team ``Pristine''.
	
	This work has made use of data from the European Space Agency (ESA) mission {\it Gaia} (\url{https://www.cosmos.esa.int/gaia}), processed by the {\it Gaia} Data Processing and Analysis Consortium (DPAC, \url{https://www.cosmos.esa.int/web/gaia/dpac/consortium}). Funding for the DPAC has been provided by national institutions, in particular the institutions participating in the {\it Gaia} Multilateral Agreement. 

	Funding for the Sloan Digital Sky Survey IV has been provided by the Alfred P. Sloan Foundation, the U.S. Department of Energy Office of Science, and the Participating Institutions. SDSS acknowledges support and resources from the Center for High-Performance Computing at the University of Utah. The SDSS web site is www.sdss.org.

    SDSS is managed by the Astrophysical Research Consortium for the Participating Institutions of the SDSS Collaboration including the Brazilian Participation Group, the Carnegie Institution for Science, Carnegie Mellon University, Center for Astrophysics | Harvard \& Smithsonian (CfA), the Chilean Participation Group, the French Participation Group, Instituto de Astrofísica de Canarias, The Johns Hopkins University, Kavli Institute for the Physics and Mathematics of the Universe (IPMU) / University of Tokyo, the Korean Participation Group, Lawrence Berkeley National Laboratory, Leibniz Institut für Astrophysik Potsdam (AIP), Max-Planck-Institut für Astronomie (MPIA Heidelberg), Max-Planck-Institut für Astrophysik (MPA Garching), Max-Planck-Institut für Extraterrestrische Physik (MPE), National Astronomical Observatories of China, New Mexico State University, New York University, University of Notre Dame, Observatório Nacional / MCTI, The Ohio State University, Pennsylvania State University, Shanghai Astronomical Observatory, United Kingdom Participation Group, Universidad Nacional Autónoma de México, University of Arizona, University of Colorado Boulder, University of Oxford, University of Portsmouth, University of Utah, University of Virginia, University of Washington, University of Wisconsin, Vanderbilt University, and Yale University.

\section*{Data Availability}
The data underlying this article will be shared on reasonable request to the corresponding author.
%%%%%%%%%%%%%%%%%%%% REFERENCES %%%%%%%%%%%%%%%%%%

\bibliographystyle{mnras}
\bibliography{PristineSgr.bib}   % bibtex database

%%%%%%%%%%%%%%%%%%%%%%%%%%%%%%%%%%%%%%

\appendix
\appendix
\section{The effect of alpha abundances on the CCD}\label{sec:alpha}

\begin{figure*}
\includegraphics[width=0.6\hsize]{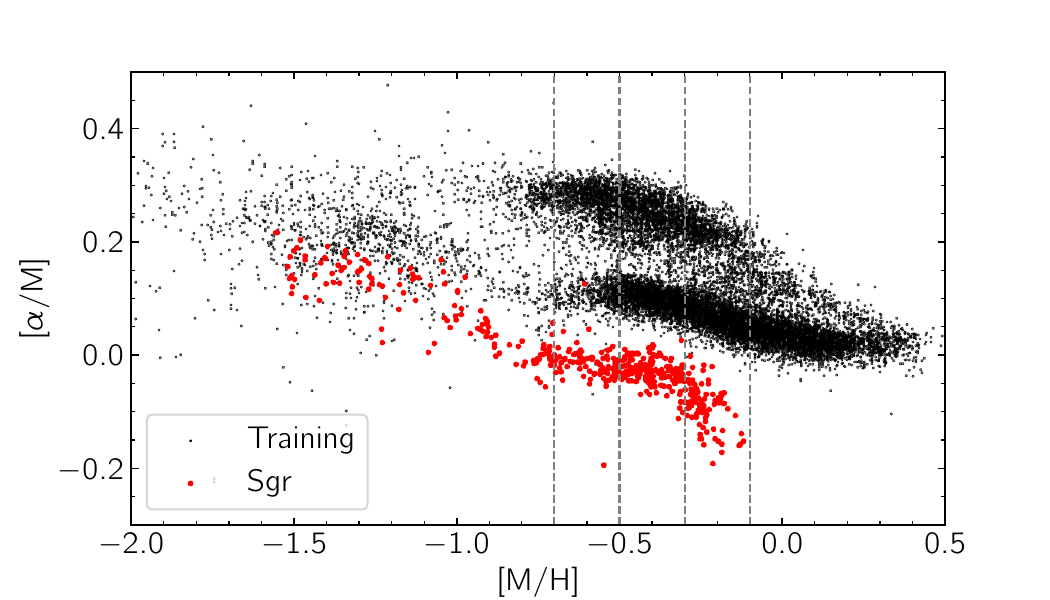}
\includegraphics[width=1.0\hsize,trim={1.0cm 0.0cm 1.0cm 0.0cm} ]{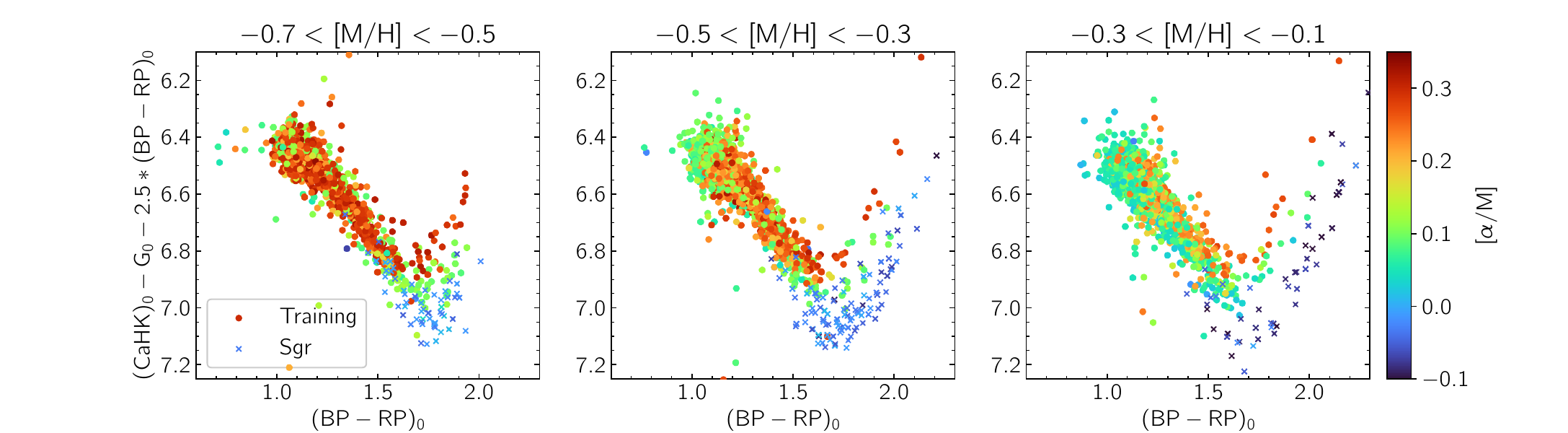}
\caption{Top: Distribution of [M/H] -- [$\alpha$/M] among APOGEE stars in the \Pristine training sample and Sgr. Bottom: \Pristine colour-colour diagram for three different metallicity ranges (see titles and grey lines in the top panel), colour-coded by [$\alpha$/M]. For $\bprp > 1.5$, three different [$\alpha$/M] sequences can clearly be seen.}
\label{fig:alpha} 
\end{figure*}

The \Pristine colour-colour diagram has been shown to be metallicity-sensitive \citep{2017MNRAS.471.2587S}, but while writing the current work, we discovered that there is also an effect due to the alpha abundances, especially for cooler stars. For this investigation, we used APOGEE abundances for giant stars in the main \Pristine survey and in Sgr, with ASPCAPFLAG~=~0 and uncertainties on [$\alpha$/M]~$< 0.1$~dex. In the top panel of Figure~\ref{fig:alpha} we present the different [M/H] vs. [$\alpha$/M] sequences between Sgr (red) and the training sample (black). The latter is split into two sequences at higher metallicity, representing thin and thick disk stars. 

We make three different metallicity bin selections, indicated by the grey vertical lines, and plot their colour-colour diagrams in the bottom panels of Figure~\ref{fig:alpha}, colour-coded by their alpha abundances. For $\bprp > 1.5$ there are clearly three different [$\alpha$/M] sequences in each of these colour-colour diagrams. For $\bprp < 1.5$ the situation is less clear. In the two more metal-poor bins, the thin and thick disk stars overlap, but in the most metal-rich bin there still appears to be some separation between them. Unfortunately there are practically no Sgr stars in APOGEE with $\bprp < 1.5$, so it is unclear whether they would overlap with the thin and thick disk stars in any of the metallicity bins. 

Although calcium is an alpha element, the Ca II H\&K lines closely trace the metallicity of most stars, down to the extremely metal-poor regime \citep{2017MNRAS.471.2587S}. The stars with $\bprp > 1.5$ have temperatures in the range of 3800~K -- 4200~K. We conclude that for giant stars of these temperatures the metallicity-sensitivity of the Ca II H\&K lines apparently breaks down -- they start to trace the alpha abundance instead.

% Don't change these lines
\bsp	% typesetting comment
\label{lastpage}

\end{document}